\RequirePackage{ifpdf}
\documentclass[12pt]{JHEP3}         % For preprint
\usepackage{amssymb,amsfonts}
\usepackage{cite}
\usepackage{amsmath}
\usepackage{graphicx}
\usepackage{MnSymbol}
\def\rmd{{\rm d}}

\newcommand{\nn}{\nonumber \\}

\newcommand{\ba}{\begin{eqnarray}}
\newcommand{\ea}{\end{eqnarray}}
\newcommand{\be}{\begin{equation}}
\newcommand{\ee}{\end{equation}}
\newcommand{\bal}{\begin{align}}
\newcommand{\eal}{\end{align}}
\newcommand{\bay}[1]{\left(\begin{array}{#1}}
\newcommand{\eay}{\end{array}\right)}

\newcommand{\Tr}{\mbox{Tr}}
\preprint{TAUP-3004/15}
\title{\center Supersymmetric R\'enyi Entropy and\\ Weyl Anomalies in\\ Six-Dimensional $(2,0)$ Theories}

\author{Yang Zhou\\

School of Physics and Astronomy, Tel-Aviv University\\ Ramat-Aviv 69978, Israel\\

{\tt E-mails : yangzhou@post.tau.ac.il}
}

\abstract{We propose a closed formula of the universal part of supersymmetric R\'enyi entropy $S_q$ for $(2,0)$ superconformal theories in six-dimensions. We show that $S_q$ across a spherical entangling surface is a cubic polynomial of $\gamma:=1/q$, with all coefficients expressed in terms of the newly discovered Weyl anomalies $a$ and $c$. This is equivalent to a similar statement of the supersymmetric free energy on conic (or squashed) six-sphere. We first obtain the closed formula by promoting the free tensor multiplet result and then provide an independent derivation by assuming that $S_q$ can be written as a linear combination of 't Hooft anomaly coefficients. We discuss a possible lower bound ${a\over c}\geq {3\over 7}$ implied by our result.}

\begin{document}

\pagestyle{plain} \setcounter{page}{1}
\newcounter{bean}
\baselineskip16pt

%\numberwithin{equation}{section}
%\setlength\parindent{1em}
%\setcounter{MaxMatrixCols}{20}

\section{Introduction}
Exact results in interacting quantum field theories are rare. Even less is known about the six-dimensional $(2,0)$ theories, although they are the local conformal field theories (CFTs) with maximal supersymmetry in the maximum number of dimensions~\cite{Nahm:1977tg,Cordova}, which actually play important roles in understanding lower dimensional supersymmetric physics~\cite{Gaiotto:2009we,Gaiotto:2009hg,Dimofte:2011ju,Bah:2012dg,Gadde:2013sca}. The main obstacle is that the proper formulation of the interacting theories is still lacking, for instance in the path integral formalism.\footnote{For the attempts to write down a Lagrangian, see for instance~\cite{Lambert:2010wm,Ho:2011ni,Chu:2012um,Bonetti:2012st,Samtleben:2012fb} and for other field theoretical attempts, see~\cite{Aharony:1997th,Aharony:1997an,ArkaniHamed:2001ie,Douglas:2010iu,Lambert:2010iw}.} This also makes it challenging to study the theories in curved spaces. In particular it is unclear how to perform the supersymmetric localization~\cite{Witten:1988ze,Nekrasov:2002qd,Pestun:2007rz} directly. 

Recently alternative approaches to $6d$ $(2,0)$ theories, such as effective actions on the moduli space and the superconformal bootstrap, are advocated in~\cite{Maxfield:2012aw,Cordova:2015vwa} and in~\cite{Beem:2014kka,Beem:2015aoa}, respectively. In particular, the Weyl anomaly coefficients $a_\mathfrak{g}$ and $c_\mathfrak{g}$ have been determined for the $(2,0)$ superconformal field theory (SCFT) characterized by a Lie algebra $\mathfrak{g}$,\footnote{$\mathfrak{g}=\mathfrak{u}(1)$ corresponds to a free Abelian tensor multiplet.}
\be\label{normalac}
\bar a_\mathfrak{g} := {a_\mathfrak{g}\over a_{\mathfrak{u}(1)}} = {16\over 7}h^\vee_\mathfrak{g} d_\mathfrak{g} + r_\mathfrak{g}\ ,\quad \bar c_\mathfrak{g} :={c_\mathfrak{g}\over c_{\mathfrak{u}(1)}} = 4 h^\vee_\mathfrak{g} d_\mathfrak{g} + r_\mathfrak{g}\ ,
\ee
where $r_\mathfrak{g}$, $d_\mathfrak{g}$ and $h^\vee_\mathfrak{g}$ are the rank, dimension and dual Coxeter number of the compact simply-laced Lie algebra $\mathfrak{g}$, respectively. $a$ and $c$ appear generally as coefficients of the anomalous trace of the stress tensor in a six-dimensional curved background~\cite{Deser:1976yx,Deser:1993yx},
\be\label{anomalousT}
\langle T_\mu^{\,\mu} \rangle \sim a\, E_6 + \sum_{i=1}^3 c_i\, I_i\ ,
\ee 
where $E_6$ is the Euler density while $I_i$ are Weyl invariants. In the presence of $(2,0)$ superconformal symmetry, $c_{i=1,2,3}$ are proportional to a single coefficient $c$. One interesting fact is that both $\bar a_\mathfrak{g}$ and $\bar c_\mathfrak{g}$ will be uniquely fixed once we assume that they are linear combinations of the 't Hooft anomaly coefficients, $h^\vee_\mathfrak{g} d_\mathfrak{g}$ and $r_\mathfrak{g}$. This can be done by combining the large $N$ values (\,from holography~\cite{Maldacena:1997re,Witten:1998xy,Aharony:1998rm,Henningson:1998gx}\,) and the free tensor multiplet values~\cite{Bastianelli:2000hi,Bastianelli:1999ab}. 

As robust observables, the 't Hooft anomalies of the continuous global symmetries in $6d$ $(2,0)$ theories have been worked out~\cite{Duff:1995wd,Witten:1996hc,Freed:1998tg,Harvey:1998bx,Yi:2001bz,Intriligator:2000eq,Ohmori:2014kda}. They are organized in an $8$-form anomaly polynomial,
\be
{\cal I}_8 = h^\vee_{\mathfrak{g}}d_{\mathfrak{g}}\, {p_2(R)\over 24} +  r_\mathfrak{g}\, {\cal I}_{\mathfrak{u}(1)}\ ,
\ee where $p_2(R)$ is the second Pontryagin class of the field strength of the $SO(5)$ R-symmetry background and ${\cal I}_{\mathfrak{u}(1)}$ is the anomaly polynomial of a free Abelian tensor multiplet.

As in other even dimensions, it is known that $a_\mathfrak{g}$ determines both the universal part\footnote{By ``universal'' we mean scheme-independent.} of the sphere partition function and the universal entanglement entropy associated with a spherical entangling surface (in flat space)~\cite{Casini:2011kv}. On the other hand, it was pointed out that $c_\mathfrak{g}$ determines both the 2-point and the 3-point functions of the stress tensor in the vacuum in flat space~\cite{Beem:2014kka,Beem:2015aoa}. Due to the intrinsic relations between the flat space stress tensor correlators and the nearly-round sphere partition function, it is therefore attempting to ask whether one can fully determine the partition function on a branched ($q$-deformed) sphere,\footnote{A branched sphere is a sphere with a conical singularity with the deformation parameter $q-1$.} which is directly related to the supersymmetric R\'enyi entropy $S_q$.

The concept supersymmetric R\'enyi entropy was first introduced in three-dimensions~\cite{Nishioka:2013haa,Huang:2014gca,Nishioka:2014mwa}, and later studied in four-dimensions~\cite{Huang:2014pda,Zhou:2015cpa,Crossley:2014oea}, five-dimensions~\cite{Alday:2014fsa,Hama:2014iea} and for free tensor multiplets in six-dimensions~\cite{Nian:2015xky}.\footnote{The supersymmetric R\'enyi entropy was recently studied in two-dimensional $(2,2)$ SCFTs~\cite{Giveon:2015cgs} in a slightly different way.} By turning on certain R-symmetry background fields (chemical potentials), one can calculate the partition function $Z_q$ on a $q$-branched sphere $\Bbb{S}_q^d$, and define the supersymmetric R\'enyi entropy as
\be\label{SREdefine}
S_q = {1\over 1-q}\left[\log Z_q(\mu(q))-q\log Z_1(0)\right]\ ,
\ee which is a supersymmetric refinement of the ordinary R\'enyi entropy (which is non-supersymmetric because of the conical singularity).\footnote{For CFTs, the R\'enyi entropy (or supersymmetric one) associated with a spherical entangling surface in flat space can be mapped to that on a sphere. Throughout this work we take the ``regularized cone'' boundary conditions.} The quantities defined in (\ref{SREdefine}) are UV divergent in general but one can extract universal parts free of ambiguities. For instance, for ${\cal N}=4$ SYM in four-dimensions, the log coefficient of $S_q$ as a function of $q$ and three chemical potentials $\mu_1,\mu_2,\mu_3$ (corresponding to three independent R-symmetry Cartans) has been shown to be protected from the interactions~\cite{Huang:2014pda}. It also receives a precise check from the holographic computation on the $5d$ BPS STU topological black holes~\cite{Huang:2014pda}. Furthermore, there are universal relations between the Weyl anomaly coefficients $a,c$ and the supersymmetric R\'enyi entropy in $4d$ ${\cal N}=1,2$ SCFTs, which provides a new way to understand the Hofman-Maldacena bounds~\cite{Zhou:2015cpa}.\footnote{Some of $a/c$ bounds by Hofman and Maldacena~\cite{Hofman:2008ar} coincide with R\'enyi entropy inequalities.} The above facts indicate that the supersymmetric R\'enyi entropy may be used as a new robust observable to understand SCFTs. 

In this work we show that the supersymmetric R\'enyi entropy of $6d$ $(2,0)$ SCFTs characterized by simply-laced Lie algebra $\mathfrak{g}$ is given by a cubic polynomial of $\gamma:={1\over q}$
\be\label{closedF}
S^{(2,0)}_\gamma = \sum_{n=0}^3 s_n (\gamma-1)^n\ ,
\ee with four coefficients
\be\label{closeds}
s_0 = {7\over 12}\,\bar a_\mathfrak{g}\ ,\quad s_1 = {1+2r_1r_2\over 12}\,\bar c_\mathfrak{g}\ ,\quad s_2 = {r_1r_2\over 12}\,\bar c_\mathfrak{g}\ ,\quad s_3 = {r_1^2r_2^2\over 12}\,{7 \bar a_\mathfrak{g}-3 \bar c_\mathfrak{g}\over 4}\ ,
\ee where $r_1$ and $r_2$ are background parameters denoting the weights of the two $U(1)$ chemical potentials associated to the two R-symmetry Cartans, satisfying the supersymmetry constraint $r_1+r_2=1$. \footnote{We only consider non-negative weights of the chemical potentials, $r_1\geq 0$ and $r_2\geq 0$.} Since both $\bar a_\mathfrak{g}$ and $\bar c_\mathfrak{g}$ are linear combinations of the 't Hooft anomaly coefficients, one may rewrite the closed formula $S^{(2,0)}_\gamma$ (\ref{closedF}) also as a linear combination of $h^\vee_\mathfrak{g} d_\mathfrak{g}$ and $r_\mathfrak{g}$,
\be\label{alterF}
S^{(2,0)}_\gamma = h^\vee_\mathfrak{g} d_\mathfrak{g}\, H_\gamma + r_\mathfrak{g}\, T_\gamma\ ,
\ee with coefficients as cubic polynomials of $\gamma$
\ba
T_\gamma = \frac{r_1^2 r_2^2}{12} (\gamma -1)^3+\frac{r_1r_2}{12} (\gamma -1)^2+\frac{1+2r_1r_2}{12} (\gamma -1)+\frac{7}{12}\ ,\\
H_\gamma = \frac{r_1^2 r_2^2}{12} (\gamma -1)^3+\frac{r_1r_2}{3} (\gamma -1)^2+\frac{1+2r_1r_2}{3} (\gamma -1)+\frac{4}{3}\ .
\ea

We derive the closed formula (\ref{closedF}) by promoting the supersymmetric R\'enyi entropy of a free tensor multiplet. The free tensor multiplet result is nothing but $T_\gamma$ in the alternative expression (\ref{alterF}), which can be directly computed using heat kernel method. Demanding that entanglement entropy $S_{\gamma=1}$ is proportional to $a_\mathfrak{g}$ and both the first and the second $\gamma$-derivatives at $\gamma=1$ are proportional to $c_\mathfrak{g}$, we could fix the coefficients $s_0, s_1$ and $s_2$ in (\ref{closedF}). We fix the remaining $s_3$ by demonstrating a precise relation between the large $\gamma$ (small $q$) behavior of supersymmetric R\'enyi entropy and the supersymmetric Casimir energy on extremely squashed sphere.\footnote{This relation was first advertised in~\cite{Zhou:2015cpa} in four-dimensions.} As a nontrivial test of our result (\ref{closedF}), we show that $H_\gamma$ precisely agrees with the holographic result computed from the BPS topological black hole in $7d$ gauged supergravity.

The fact that $T_\gamma$ and $H_\gamma$ are precisely the free multiplet result and the holographic result, respectively, actually provides an alternative derivation of the closed formula (\ref{closedF}). Consider $T_\gamma$ and $H_\gamma$ as independent results from the free field computation and the holographic dual, respectively, one can uniquely determine the closed formula of $S^{(2,0)}_\gamma$ by assuming that $S^{(2,0)}_\gamma$ is a linear combination of the 't Hooft anomaly coefficients. This assumption can be reasonably imposed once we are aware of any two of $s_{n=0,1,2,3}$ as linear combinations of $h^\vee_\mathfrak{g} d_\mathfrak{g}$ and $r_\mathfrak{g}$.

This paper is organized as follows. We begin with the general study of the relations between the perturbative supersymmetric R\'enyi entropy around $q=1$ and the integrated correlation functions (stress tensor and R-current) in Section \ref{Pexp}, which works for general dimensions. We focus on the first and the second derivative at $q=1$. Then we review the supersymmetric R\'enyi entropy of free tensor multiplets in Section \ref{SREtensor}. Built up on these facts, we propose a way to determine the supersymmetric R\'enyi entropy for interacting $(2,0)$ theories in Section \ref{SREint}. In Section \ref{SREasympt}, we show a general relation between the $q\to 0$ behavior of supersymmetric R\'enyi entropy and supersymmetric Casimir energy, which is used to determine the remaining unfixed coefficient in the proposed formula in the previous section. Finally we give a precise test of our results by comparing with the holographic results in Section \ref{SREholography}.

\section{Near $q=1$ expansion}
\label{Pexp}
We begin with the perturbative expansion of supersymmetric R\'enyi entropy (associated with spherical entangling surface) around $q=1$. This can be considered as an extension of the previous study of the ordinary R\'enyi entropy near $q=1$. Although our main concern will be $6d$ $(2,0)$ SCFTs, we keep the discussions in this section valid for any SCFT with conserved R-symmetries in $d$-dimensions. 

Following the way in~\cite{Perlmutter:2013gua,Lee:2014zaa} \footnote{See~\cite{Hung:2014npa} from the viewpoint of twisted operator.}, we consider the supersymmetric partition function on $\Bbb{S}^1_q\times\Bbb{H}^{d-1}$ with background gauge fields (R-symmetry chemical potentials), which can be used to compute the supersymmetric R\'enyi entropy across a spherical entangling surface, see $\Bbb{S}^{d-2}$, in flat space.
We work in grand canonical ensemble. The partition function on $\Bbb{S}^1_{\beta=2\pi q}\times\Bbb{H}^{d-1}$ can be written as
\be
Z[\beta,\mu] = \Tr \left(e^{-\beta(\hat E-\mu \hat Q)}\right)\ .
\ee
The state variables can be computed as follows
\ba
E ~&=&~ \left({\partial I\over \partial\beta}\right)_\mu~ - {\mu\over\beta}\left({\partial I\over \partial\mu}\right)_\beta\ ,\label{Evariable}\\
S ~&=&~ \beta \left({\partial I\over \partial\beta}\right)_\mu~ - I\ ,\\
Q ~&=&~ -{1\over\beta}\left({\partial I\over \partial\mu}\right)_\beta\ ,\label{Qvariable}
\ea where $I:=-\log Z$.
Therefore we get energy expectation value by (\ref{Evariable})
\be\label{Eexp}
E={\Tr \left(e^{-\beta(\hat E-\mu \hat Q)}\,\hat E\right) \over \Tr \left(e^{-\beta(\hat E-\mu \hat Q)}\right)}\ ,
\ee and the charge expectation value by (\ref{Qvariable})
\be\label{Qexp}
Q={\Tr \left(e^{-\beta(\hat E-\mu \hat Q)}\,\hat Q\right) \over \Tr \left(e^{-\beta(\hat E-\mu \hat Q)}\right)}\ .
\ee
In the presence of supersymmetry, both inverse temperature $\beta$ and chemical potential $\mu$ are functions of a single variable $q$ therefore $I$ is considered as 
\be I_q:=I[\beta(q),\mu(q)]\ .
\ee The supersymmetric R\'enyi entropy is defined as
\be
S_q = {q I_1- I_q\over 1-q}\ .
\ee
Consider the Taylor expansion around $q=1$, with $\delta q=q-1$,
\be\label{renyiexpansion}
S_q = S_{\text{EE}} + \sum_{n=2}^\infty{1\over n!}\left({\partial^n I_q\over\partial q^n}\right)_{q=1}\delta q^{n-1}\ .
\ee

\subsection{$\partial_q I_q$}
We will first consider $\partial_q I_q$. The first derivative with respect to $q$ can be written as
\be
{d I_q\over d q} = \left({\partial I\over \partial\beta}\right)_\mu\, \beta^\prime(q) + \left({\partial I\over \partial\mu}\right)_\beta\, \mu^\prime(q)\ .
\ee
Using (\ref{Evariable}) and (\ref{Qvariable}), we can rewrite it as
\be
{d I_q\over d q} = (E-\mu Q)\, \beta^\prime(q) - \beta Q\, \mu^\prime(q)\ .
\ee
The $q$-dependence of the temperature and the chemical potential can be read off from the supersymmetric background (including metric and R-symmetry gauge field),
\be\label{qbackground}
\beta(q) = 2\pi q\ ,\quad \mu(q)=\alpha{q-1\over q}\ ,
\ee where $\beta(q)$ is determined by the geometric fact and $\mu(q)$ is solved from the Killing spinor equation on the background. $\alpha$ is some number which may be different in various rigid supersymmetric backgrounds.\footnote{In the case of multiple chemical potentials, one should use $\alpha_{i=1,2...R}$, where $R$ denotes the number of $U(1)$ R-symmetry Cartans. $i$ should be summed over for $\alpha_i Q^i$. } The first $q$-derivative of $I_q$ is simplified by using (\ref{qbackground}) 
\be\label{firstD}
I^\prime_q = 2\pi(E-\alpha Q)\ .
\ee
Notice that in general both $E$ and $Q$ are functions of $q$. Also $E$ and $Q$ here are expectation values rather than operators.

\subsection{$S^\prime_{q=1}$ and $I^{\prime\prime}_{q=1}$}
From (\ref{renyiexpansion}) we see that
\be
S^\prime_{q=1} = {1\over 2} I^{\prime\prime}_{q=1}\ .
\ee
Let us take one more derivative above on the first derivative (\ref{firstD}) and take use of (\ref{Eexp}) and (\ref{Qexp})
\be
I^{\prime\prime}_q = -4\pi^2\left( {\Tr \left(e^{-\beta(\hat E-\mu \hat Q)}\,(\hat E-\alpha \hat Q)^2\right) \over \Tr \left(e^{-\beta(\hat E-\mu \hat Q)}\right)}-{\left[\Tr \left(e^{-\beta(\hat E-\mu \hat Q)}\,(\hat E-\alpha \hat Q)\right) \right]^2\over \left[\Tr \left(e^{-\beta(\hat E-\mu \hat Q)}\right)\right]^2}\right)\ ,\label{sreprimemu}
\ee
which can be simplified in the limit $q\to 1$ by using $\mu=0$ at $q=1$
\be
S^\prime_{q=1} = -2\pi^2 \left({\Tr \left(e^{-\beta \hat E}\,(\hat E-\alpha \hat Q)^2\right) \over \Tr \left(e^{-\beta \hat E}\right)}-{\left[\Tr \left(e^{-\beta \hat E}\,(\hat E-\alpha \hat Q)\right) \right]^2\over \left[\Tr \left(e^{-\beta \hat E}\right)\right]^2}\right)_{q=1}\ .
\ee
This can be rewritten as connected correlators
\be
S^\prime_{q=1} = -2\pi^2 \left[\langle \hat E \hat E\rangle^c+\alpha^2\langle \hat Q \hat Q\rangle^c-2\alpha\langle \hat E \hat Q\rangle^c\right]_{\Bbb{S}^1_{q=1}\times\Bbb{H}^{d-1}}\ ,
\ee where we have used $[\hat E,\hat Q]=0$ to flip the order of $\hat E$ and $\hat Q$. Given that $\langle \hat E \hat Q\rangle^c=0$ and $\langle \hat E \hat E\rangle^c$ has been computed in~\cite{Perlmutter:2013gua}, we get
\be\label{sreprime0}
S^\prime_{q=1} =-V_{d-1}{\pi^{d/2+1}\Gamma(d/2)(d-1)\over (d+1)!}C_T -2\pi^2 \alpha^2\int _{\Bbb{H}^{d-1}}\int _{\Bbb{H}^{d-1}}\langle J_\tau(x)J_\tau(y)\rangle^c_{q=1}\ .
\ee $C_T$ is defined in the flat space correlator
\be
\langle T_{ab}(x)T_{cd}(0)\rangle = {C_T\over x^{2d}}I_{ab,cd}(x)\ ,
\ee where
\be
I_{ab,cd}(x) = {1\over 2}\left(I_{ac}(x)I_{bd}(x)+ I_{ad}(x)I_{bc}(x)\right)-{1\over d}\delta_{ab}\delta_{cd}\ ,\quad I_{ab}(x)=\delta_{ab}-2{x_ax_b\over x^2}\ .
\ee
Now the task is to compute the second term in (\ref{sreprime0}). Following the way of computing $\langle TT\rangle$ on the hyperbolic space $\Bbb{S}^1_{q=1}\times\Bbb{H}^{d-1}$, one can take use of the flat space correlators in the CFT vacuum. The result is~\footnote{$\langle J\hat Q\rangle$ was first computed in~\cite{Belin:2013uta}.}
\be
\langle \hat Q \hat Q\rangle^c = -{\pi^{d-1\over 2}V_{d-1}\over 2^{d-2}(d-1)\Gamma({d-1\over 2})}C_v\ ,
\ee where $C_v$ is defined in the current correlator in flat space
\be
\langle J_a(x)J_b(0) \rangle = {C_v\over x^{2(d-1)}} I_{ab}(x)\ .
\ee
Then our final result of $S^\prime_{q=1}$ becomes
\be\label{susyprimeG}
S^\prime_{q=1}=-V_{d-1}\left({\pi^{{d\over 2}+1}\Gamma({d\over 2})(d-1)\over (d+1)!}C_T -\alpha^2{\pi^{d+3\over 2}\over 2^{d-3}(d-1)\Gamma({d-1\over 2})}C_v\right)\ ,
\ee
which tells us that the first $q$-derivative of supersymmetric R\'enyi entropy at $q=1$ is given by a linear combination of $C_T$ and $C_v$.\footnote{In another word, a linear combination of the integrated stress tensor 2-point function and the integrated R-current 2-point function.} This is intuitively expected because in the presence of supersymmetry, taking the derivative with respect to $q$ is equivalent to taking the derivative with respect to $g_{\tau\tau}$ and $A_\tau$ in the same time.\footnote{This was first suggested in~\cite{Huang:2014pda}.} $q$-deformation can be often equivalent to the squashing $b:=\sqrt{q}$, therefore this formula also shows the relation between $\partial^2_{b=1}$ of the free energy on squashed sphere and flat space correlators.  It is clear from the above derivation that this formula works both for free theories and interacting SCFTs in general $d$-dimensions. In the particular case of $6d$ $(2,0)$ SCFTs, the 2-point function of the stress tensor is determined by the central charge $c_\mathfrak{g}$ in (\ref{normalac})~\cite{Beem:2014kka,Beem:2015aoa}. Therefore the integrated 2-point function is proportional to $c_\mathfrak{g}$. Furthermore, $S^\prime_{q=1}$ is also proportional to $c_\mathfrak{g}$, because the stress tensor and the R-current in the right hand side of (\ref{susyprimeG}) live in the same supermultiplet.\footnote{For $(2,0)$ tensor multiplet, this supermultiplet was studied explicitly in~\cite{Bergshoeff:1999db}.} The same thing happens in ${\cal N}=4$ SYM~\cite{Huang:2014pda}.

\subsection{$S^{\prime\prime}_{q=1}$ and $I^{\prime\prime\prime}_{q=1}$}
From (\ref{renyiexpansion}) we see that
\be
S^{\prime\prime}_{q=1} = {1\over 6} I^{\prime\prime\prime}_{q=1}\ .
\ee
One may go straightforward to compute $I^{\prime\prime\prime}_{q}$ by taking one more derivative above on (\ref{sreprimemu})
\ba
{I^{\prime\prime\prime}_q\over 8\pi^3} &=& {\Tr \left(e^{-\beta(\hat E-\mu \hat Q)}\,(\hat E-\alpha \hat Q)^3\right) \over \Tr \left(e^{-\beta(\hat E-\mu \hat Q)}\right)}-3{\Tr \left(e^{-\beta(\hat E-\mu \hat Q)}\,(\hat E-\alpha \hat Q)^2\right)\Tr\left(e^{-\beta(\hat E-\mu \hat Q)}\,(\hat E-\alpha \hat Q)\right)\over \left[\Tr \left(e^{-\beta(\hat E-\mu \hat Q)}\right)\right]^2}\nn
&&~~+2{\left[\Tr \left(e^{-\beta(\hat E-\mu \hat Q)}\,(\hat E-\alpha \hat Q)\right) \right]^3\over \left[\Tr \left(e^{-\beta(\hat E-\mu \hat Q)}\right)\right]^3}\ ,
\ea which may be simplified at $q=1$ where $\mu=0$
\ba
{I^{\prime\prime\prime}_{q=1}\over 8\pi^3} &=& \bigg({\Tr \left(e^{-\beta \hat E}\,(\hat E-\alpha \hat Q)^3\right) \over \Tr e^{-\beta \hat E}}-3{\Tr \left(e^{-\beta \hat E}\,(\hat E-\alpha \hat Q)^2\right)\Tr\left(e^{-\beta \hat E}\,(\hat E-\alpha \hat Q)\right)\over \left[\Tr e^{-\beta \hat E}\right]^2} \nn &+& 2{\left[\Tr (e^{-\beta \hat E}\,(\hat E-\alpha \hat Q))\right]^3\over \left[\Tr e^{-\beta \hat E}\right]^3}\bigg)_{q=1}\ .
\ea This can be further written in terms of connected correlation functions,
\be\label{3pointc}
S^{\prime\prime}_{q=1} ={1\over 6} I^{\prime\prime\prime}_{q=1}={4\pi^3\over 3}\left[\langle \hat E\hat E\hat E\rangle^c-\alpha^3\langle \hat Q\hat Q\hat Q\rangle^c-3\alpha\langle \hat E\hat E\hat Q\rangle^c+3\alpha^2\langle \hat E\hat Q\hat Q\rangle^c\right]_{\Bbb{S}^1_{q=1}\times\Bbb{H}^{d-1}}\ ,
\ee where we have used $[\hat E,\hat Q]=0$ because $\hat Q$ is conserved charge. The integrated correlators in (\ref{3pointc}) can be computed by transforming the corresponding flat space correlators, $\langle TTT\rangle, \langle JJJ\rangle, \langle TTJ\rangle, \langle TJJ\rangle$ in the CFT vacuum.\footnote{We leave the explicit computations of these correlators elsewhere.} These correlators in flat space can be determined up to some coefficients for general CFTs in $d$-dimensions by conformal Wald identities~\cite{Osborn:1993cr,Erdmenger:1996yc}. In the presence of $6d$ $(2,0)$ superconformal symmetry, both the 2- and 3-point functions of the stress tensor supermultiplet are uniquely determined in terms of a single parameter, the central charge $c_\mathfrak{g}$~\cite{Beem:2014kka,Beem:2015aoa}. And the right hand side of (\ref{3pointc}) should be proportional to $c_\mathfrak{g}$, because the stress tensor and the R-current belong to the same supermultiplet.\footnote{By representation theory, the stress tensor belongs to a half BPS multiplet.  In superspace, the 2-, 3- and 4-point functions of all half BPS multiplets are known to admit a unique structure~\cite{Eden:2001wg,Arutyunov:2002ff,Dolan:2004mu}.} The same thing can be seen in ${\cal N}=4$ SYM~\cite{Huang:2014pda}.

\section{Abelian tensor multiplet}
\label{SREtensor}
The six-dimensional $(2,0)$ superconformal algebra is $\mathfrak{osp}(8^*|4)$. While it is easy to identify a free Abelian tensor multiplet that realizes the $(2,0)$ superconformal symmetry, the existence of interacting $(2,0)$ theories was only inferred from decoupling limits of string constructions~\cite{Witten:1995zh,Strominger:1995ac,Witten:1995em}. See for instance~\cite{MooreLecute} for a review of various aspects of $6d$ $(2,0)$ theories.

Now we review the supersymmetric R\'enyi entropy of free tensor multiplets~\cite{Nian:2015xky}. For free fields, the R\'enyi entropy associated with a spherical entangling surface in flat space can be computed by working on a hyperbolic space $\Bbb{S}^1_\beta\times\Bbb{H}^5$ and using heat kernel method.\footnote{Six-dimensional $(2,0)$ theories have been studied in $AdS_5\times S^1$ recently in the viewpoint of rigid holography~\cite{Aharony:2015zea}.} A six-dimensional $(2,0)$ tensor multiplet includes 5 real scalars, 2 Weyl fermions and a 2-form field with self-dual strength. The 2-form field with self-dual strength can be considered as a chiral 2-form field with half of the degrees of freedom.

\subsection{Heat kernel}
The partition function of free fields on $\Bbb{S}^1_{\beta=2\pi q}\times\Bbb{H}^5$ can be obtained by heat kernel method,\footnote{For R\'enyi entropy of free fields in other higher dimensions, see for instance~\cite{Casini:2010kt,Klebanov:2011uf,Fursaev:2012mp,Dowker:2012rp}.}
\be\label{partitionkernel}
\log Z(\beta) = {1\over 2} \int_0^\infty {dt\over t}K_{\mathbb{S}^1_\beta\times \mathbb{H}^5}(t)\ ,
\ee where $K_{\mathbb{S}^1_\beta\times \mathbb{H}^5}(t)$ is the heat kernel of the associated conformal Laplacian. The kernel can be factorized when the spacetime is a direct product,
\be
K_{\mathbb{S}^1_\beta\times \mathbb{H}^5}(t)=K_{\mathbb{S}^1_\beta}(t)\, K_{\mathbb{H}^5}(t)\ .
\ee The kernel on a circle $K_{\mathbb{S}^1_\beta}(t)$ is known to be~\footnote{For fermions, the boundary conditions are anti-periodic.}
\be
K_{\mathbb{S}^1_\beta}(t)={\beta\over \sqrt{4\pi t}}\sum_{n\neq 0,\in\mathbb{Z}} e^{-\beta^2n^2\over 4t}\ .
\ee
In the presence of a chemical potential $\mu$, it is twisted to be~\cite{Belin:2013uta}
\be\label{twkernel}
\widetilde K_{\mathbb{S}^1_\beta}(t)={\beta\over \sqrt{4\pi t}}\sum_{n\neq 0,\in\mathbb{Z}} e^{{-\beta^2n^2\over 4t}+i2\pi n \mu+i\pi n f}\ ,
\ee where $f=0$ for scalars and $f=1$ for fermions. Finally the kernels on the hyperbolic space $K_{\mathbb{H}^5}(t)$ can be written as follows because $\mathbb{H}^5$ is homogeneous,
\be
K_{\mathbb{H}^5}(t) = \int d^5x \sqrt{g}~K_{\mathbb{H}^5}(x,x,t)=V_5\, K_{\mathbb{H}^5}(0,t)\ .
\ee The regularized volume $V_5=\pi^2\log(\ell/\epsilon)$. $\epsilon$ is the UV cutoff of the theory in the original space \footnote{This is the $q$-fold space with a conical singularity, which is used to compute R\'enyi entropy by replica trick. } and $\ell$ is the curvature radius of $\mathbb{H}^5$. Note that the kernels $K_{\mathbb{H}^5}(0,t)$ for free fields with different spins are known. See~\cite{Nian:2015xky} and references there.

\subsection{R\'enyi entropy}
The total R\'enyi entropy of a tensor multiplet can be obtained by summing up the contributions of 5 real scalars, 2 Weyl fermions and a chiral 2-form,
\be
S^{free}_q = 5\times {S^s_q\over 2} + 2 S^f_q + {S^v_q\over 2}\ ,
\ee where the R\'enyi entropy for fields with different spins can be computed by using the corresponding heat kernels. For the details of this computation we refer to~\cite{Nian:2015xky}. We will instead list the results here. The R\'enyi entropy of a $6d$ real scalar is
\be
S^s_q = \frac{(q+1) \left(3 q^2+1\right) \left(3 q^2+2\right)}{15120 q^5}{V_5\over \pi^2}\ ,
\ee and the R\'enyi entropy of a $6d$ Weyl fermion is
\be
S^f_q = \frac{(q+1) \left(1221 q^4+276 q^2+31\right)}{120960 q^5}{V_5\over \pi^2}\ ,
\ee and that of a $6d$ 2-from field is
\be
S^v_q = \frac{(q+1) \left(37 q^2+2\right)+877 q^4+4349q^5}{5040 q^5}{V_5\over \pi^2}\ .
\ee
It is worth to mention that, to get the correct R\'enyi entropy for the two form field, one has to take into account a $q$-independent constant shift due to the edge modes~\cite{Nian:2015xky}, like what should  be done for the gauge field in $4d$~\cite{Huang:2014pfa,Eling:2013aqa}.
Finally the R\'enyi entropy for a free $(2,0)$ tensor multiplet is
\be\label{renyifree}
S^{free}_q = {(q+1)(28 q^2 + 3) + 313 q^4 + 1305 q^5\over 2880 q^5}{V_5\over \pi^2}\ .
\ee
It has been checked that $\partial^0_{q=1}$, $\partial^1_{q=1}$ and $
\partial^2_{q=1}$ of $S^{free}_q$ are consistent~\cite{Nian:2015xky} with the previous results  about the tensor multiplet~\cite{Bastianelli:2000hi,Bastianelli:1999ab,Safdi:2012sn}. 

\subsection{$S_q$ and $S_\gamma$}
Before moving on, let us represent $S^{free}_q$ in terms of 
$$S_{\gamma}:={\pi^2\over V_5}\, S_q\ ,~\text{with}~~ \gamma:=1/q\ , $$
\be\label{tensorformula}
S^{free}_{\gamma}=\frac{1}{960} (\gamma -1)^5+\frac{1}{160} (\gamma -1)^4+\frac{7}{288} (\gamma -1)^3+\frac{1}{18} (\gamma -1)^2+\frac{\gamma -1}{6}+\frac{7}{12}\ .
\ee The reason why $S_\gamma$ is convenient is that, the series expansion near $\gamma=1$ has finite terms while the expansion of $S_q$ near $q=1$ has infinite terms. We will use $S_{\gamma}$ instead of $S_q$ to express R\'enyi entropy and supersymmetric R\'enyi entropy from now on.
It is worth to note the relations between the derivatives with respect to $q$ and the derivatives with respect to $\gamma$ at $q=1/\gamma =1$,
\be\label{derivatives}
\partial_\gamma S_\gamma = -\partial_q S_q\bigg |_{q=1/\gamma=1}\cdot {\pi^2\over V_5} \ ,\quad \partial^2_\gamma S_\gamma = \left(2\partial_q S_q+\partial^2_q S_q\right) \bigg |_{q=1/\gamma=1}\cdot {\pi^2\over V_5}\ .
\ee

\subsection{Supersymmetric R\'enyi entropy}
The supersymmetric R\'enyi entropy of a free tensor multiplet can be computed by the twisted kernel (\ref{twkernel}) on the supersymmetric background. The R-symmetry group of 6$d$ $(2,0)$ theories is $SO(5)$, which has two $U(1)$ Cartans. Therefore one can turn on two independent R-symmetry background gauge fields (chemical potentials) to twist the boundary conditions for scalars and fermions along the replica circle $\Bbb{S}_\beta^1$. A general analysis of the Killing spinor equation on the conic space ($\Bbb{S}^6_q$ or $\Bbb{S}_{\beta=2\pi q}^1\times\Bbb{H}^5$) leads to the solution of the R-symmetry chemical potential~\cite{Nian:2015xky} \footnote{The Killing spinors on round sphere have been explored in~\cite{Lu:1998nu}.}
\be\label{kschemical}
\mu(q) :=k_iA^i= {q-1\over 2}\ ,
\ee with $k_1$ and $k_2$ being the R-charges of the Killing spinor under the two $U(1)$ Cartans, respectively. We choose $k_1=k_2={1\over 2}$ and the two background fields can be expressed as
\be\label{chemicalback}
A^1 = (q-1)\, r_1\ ,\quad A^2 = (q-1)\, r_2\ ,\quad \text{with}~~r_1+r_2=1\ .
\ee
This is the most general background satisfying (\ref{kschemical}). For each component field in the tensor multiplet, one has to first figure out the Cartan charges $k_1$ and $k_2$ and then compute the chemical potential by $k_1A^1+k_2A^2$. Then one can compute the free energy on $\Bbb{S}_\beta^1\times\Bbb{H}^5$ using the twisted heat kernel and get the supersymmetric R\'enyi entropy. For details, see~\cite{Nian:2015xky}. 

After summing up all the component fields, the final supersymmetric R\'enyi entropy in terms of $\gamma$ can be expressed as,\footnote{Although the form of this expression is a series expansion, the result itself is complete.} 
\be\label{freeformula}
 S^{free}_\gamma = \frac{1}{12} r_1^2 r_2^2 (\gamma -1)^3+\frac{1}{12} r_1r_2 (\gamma -1)^2+\frac{1}{12} (1+2r_1r_2) (\gamma -1)+\frac{7}{12}\ .
\ee
It is worth to note that, for a single $U(1)$ background, $r_1=1, r_2=0$, the result becomes
\be
S_\gamma = \frac{1}{12} (\gamma +6)\ ,
\ee while for two $U(1)$ backgrounds with equal values, $r_1=r_2={1\over 2}$, we have
\be
 S_\gamma = \frac{1}{192}(\gamma -1)^3+\frac{1}{48}(\gamma -1)^2+\frac{1}{8}(\gamma -1)+\frac{7}{12}\ .
\ee

\section{Interacting $(2,0)$ theories}
\label{SREint}
Having obtained the supersymmetric R\'enyi entropy (\ref{freeformula}) for a free tensor multiplet, we now try to promote it to a general form which may work for interacting $(2,0)$ SCFTs,
\be\label{ansatz}
S^{(2,0)}_\gamma = \frac{r_1^2 r_2^2}{12} \cdot A\, (\gamma -1)^3+\frac{r_1r_2}{12}\cdot B\, (\gamma -1)^2+\frac{1+2r_1r_2}{12}\cdot C\, (\gamma -1)+\frac{7}{12}D\ ,
\ee
where the coefficients $A\,,B\,,C\,,D$ will depend on the specific theory.\footnote{$S^{(2,0)}_\gamma$ should be a cubic polynomial of $\gamma$, which is the unique option compatible with both free field result and holographic result (as we will see). The same thing happens in ${\cal N}=4$ SYM. Here we see an essential difference between the ordinary R\'enyi entropy and the supersymmetric one, because the type of $q$ scaling in the ordinary R\'enyi entropy is not protected~\cite{Galante:2013wta,Lee:2014zaa}.} The factors carrying $r_1$ and $r_2$ should stay the same as that appearing in the free multiplet result (\ref{freeformula}) because they originally come from the $\alpha_i$ ($\alpha_1=r_1, \alpha_2=r_2$) in (\ref{susyprimeG})(\ref{3pointc}), which are background parameters independent of the specific theory.
Later we will see that precisely the same factors appear in the holographic supersymmetric R\'enyi entropy, which confirms this fact.
\subsection{$S^{(2,0)}_{\gamma=1}$ and $a_{\mathfrak{g}}$}
We would like to first determine the coefficient $D$ in (\ref{ansatz}). This can be done by using the fact that, the entanglement entropy associated with a spherical entangling surface, which is nothing but $S_{\gamma=1}$, is proportional to $a$, where $a$ is the $a$-type Weyl anomaly. This is true for general CFTs in even dimensions as shown in~\cite{Casini:2011kv}. Therefore
\be
{S^{(2,0)}_{\text{EE}}\over S^{free}_{\text{EE}}}={a_{\mathfrak{g}}\over a_{\mathfrak{u}(1)}}\ .
\ee
This allows us to fix
\be
D={a_{\mathfrak{g}}\over a_{\mathfrak{u}(1)}}={16\over 7}h^\vee_{\mathfrak{g}}d_{\mathfrak{g}}+r_{\mathfrak{g}}\ ,
\ee
where we have used the $a$-type Weyl anomaly result in $6d$ $(2,0)$ theories~\cite{Cordova:2015vwa}.

\subsection{$\partial S^{(2,0)}_{\gamma=1}, \partial^2S^{(2,0)}_{\gamma=1}$ and $c_{\mathfrak{g}}$}
The coefficients $C$ and $B$ in (\ref{ansatz}) are determined by the first and the second $\gamma$-derivatives of $S^{(2,0)}_\gamma$ at $\gamma=1$, respectively. $\gamma$-derivatives can be translated into $q$-derivatives. Taking $q$-derivatives can be equivalently considered as taking derivatives with respect to background fields, therefore $\partial S^{(2,0)}_{\gamma=1}$ and $\partial^2S^{(2,0)}_{\gamma=1}$ are intrinsically related to the corresponding correlators. This has been illustrated in Section \ref{Pexp}.

Explicitly, the first $\gamma$-derivative (which is minus the $q$-derivative at $q=1/\gamma=1$) is determined by a linear combination of the integrated stress tensor 2-point function and the integrated R-current 2-point function. The first $q$-derivative at $q=1$ is given by the formula (\ref{susyprimeG}),
\be
S^\prime_{q=1} = -V_{d-1}\left({\pi^{{d\over 2}+1}\Gamma({d\over 2})(d-1)\over (d+1)!}C_T -\alpha^2{\pi^{d+3\over 2}\over 2^{d-3}(d-1)\Gamma({d-1\over 2})}C_v\right)\ ,
\ee
which works for general SCFTs with conserved R-symmetries in $d$-dimensions.

Similarly the second $\gamma$-derivative at $\gamma=1$ is related to $q$-derivatives by (\ref{derivatives}). The second $q$-derivative at $q=1$ is determined by a linear combination of the integrated stress tensor 3-point function, the integrated R-current 3-point function and some mixed 3-point functions. This is given explicitly by (\ref{3pointc})
\be
S^{\prime\prime}_{q=1} ={1\over 6} I^{\prime\prime\prime}_{q=1}={4\pi^3\over 3}\left[\langle \hat E\hat E\hat E\rangle^c-\alpha^3\langle \hat Q\hat Q\hat Q\rangle^c-3\alpha\langle \hat E\hat E\hat Q\rangle^c+3\alpha^2\langle \hat E\hat Q\hat Q\rangle^c\right]_{\Bbb{S}^1_{q=1}\times\Bbb{H}^{d-1}}\ ,
\ee
which also works for general SCFTs with conserved R-symmetries in $d$-dimensions.

In the particular case of $6d$ $(2,0)$ SCFTs, all the above two- and three-point functions may be uniquely determined in terms of a single parameter, the central charge $c_\mathfrak{g}$ (\ref{normalac}), as discussed in Section \ref{Pexp}. \footnote{This actually explains the universal ratio $4N^3$ between the explicit results on $\langle TT\rangle, \langle TTT\rangle, \langle JJ\rangle, \langle JJJ\rangle$ in holography and those in free tensor multiplets~\cite{Bastianelli:1999ab, Manvelyan:2000ef}.}

Due to the above facts, the straightforward idea to get $\partial S^{(2,0)}_{\gamma=1}$ and $\partial^2S^{(2,0)}_{\gamma=1}$ for interacting theories is to multiply
\be
{c_{\mathfrak{g}}\over c_{\mathfrak{u}(1)}} = 4h^\vee_{\mathfrak{g}}d_{\mathfrak{g}} + r_{\mathfrak{g}}
\ee to the free multiplet values in (\ref{freeformula}).
This actually means we can fix
\be
B=C=4h^\vee_{\mathfrak{g}}d_{\mathfrak{g}} + r_{\mathfrak{g}}\ .
\ee
The remaining coefficient $A$ will be fixed as
\be
A=h^\vee_{\mathfrak{g}} d_{\mathfrak{g}} + r_{\mathfrak{g}}
\ee  in the next section by studying the asymptotic $q:=1/\gamma\to 0$ behavior of the supersymmetric R\'enyi entropy.
Obviously, the leading contribution in the limit $\gamma\to\infty$ is controlled only by $A$.

\subsection{A closed formula}
As a summary, we can completely determine a closed formula of supersymmetric R\'enyi entropy for $(2,0)$ SCFTs characterized by simply-laced Lie algebra $\mathfrak{g}$
\ba
S^{(2,0)}_\gamma&&=\frac{r_1^2r_2^2}{12} (h^\vee_{\mathfrak{g}} d_{\mathfrak{g}} + r_{\mathfrak{g}})\,(\gamma -1)^3 +\frac{r_1r_2}{12} (4 h^\vee_{\mathfrak{g}} d_{\mathfrak{g}} + r_{\mathfrak{g}})\,(\gamma -1)^2\nn
&&~~+\frac{1+2r_1r_2}{12} (4 h^\vee_{\mathfrak{g}} d_{\mathfrak{g}} + r_{\mathfrak{g}})\,(\gamma -1)+\left(\frac{4 h^\vee_{\mathfrak{g}} d_{\mathfrak{g}}}{3}+\frac{7 r_{\mathfrak{g}}}{12}\right)\ ,\label{SREfull}\\
=\frac{r_1^2r_2^2}{48}&&(7\bar a_{\mathfrak{g}} -3\bar c_{\mathfrak{g}})\,(\gamma -1)^3 +\frac{r_1r_2 }{12}\,\bar c_{\mathfrak{g}}\,(\gamma -1)^2+\frac{1+2r_1r_2}{12}\, \bar c_{\mathfrak{g}}\,(\gamma -1)+\frac{7}{12}\, \bar a_{\mathfrak{g}}\,,\label{fullSRE}
\ea where in the last line we have used the normalized Weyl anomalies defined in (\ref{normalac}).

For a single $U(1)$ chemical potential,
\be
r_1=1\ ,\quad r_2=0\ ,
\ee the result is simplified to be
\ba
S^{(2,0)}_\gamma &=& \frac{1}{12}\, \bar c_{\mathfrak{g}}\,(\gamma -1)+\frac{7}{12}\, \bar a_{\mathfrak{g}}\ ,\label{boundtest0}\\
&=&h^\vee_{\mathfrak{g}} d_{\mathfrak{g}}\left({1\over 3}\gamma+1\right)+r_{\mathfrak{g}}{\left(\gamma+6\right)\over 12}\ .
\ea
As for two $U(1)$ chemical potentials with equal values,
\be
r_1=r_2={1\over 2}\ ,
\ee the result is simplified to be
\ba
S^{(2,0)}_\gamma&=&\frac{1}{192\times 4}(7\bar a_{\mathfrak{g}} -3\bar c_{\mathfrak{g}})\,(\gamma -1)^3 +\frac{1 }{48}\,\bar c_{\mathfrak{g}}\,(\gamma -1)^2+\frac{1}{8}\, \bar c_{\mathfrak{g}}\,(\gamma -1)+\frac{7}{12}\, \bar a_{\mathfrak{g}}\ ,\label{boundtest}\\
&=& { 175 +67 \gamma+13 
\gamma^2+\gamma^3\over 192}h^\vee_{\mathfrak{g}} d_{\mathfrak{g}}+{91+19 \gamma+\gamma^2+\gamma^3\over 192} r_{\mathfrak{g}}\ .
\ea

\section{$q\to 0$ asymptotics}
\label{SREasympt}
In this section we discuss the $q\to 0$ limit ($\gamma\to\infty$) of supersymmetric R\'enyi entropy $S_q$. Recall the definition of $S_q$
\be
S_q={qI_1-I_q\over 1-q}\ .
\ee Assuming that in the limit $q\to 0$ the free energy behaves
\be
I_q=I_{(0)}q^{-\alpha}+\cdots\ ,
\ee where $\alpha\geq 0$, one can easily get
\be
S_{q\to 0} = -I_{q\to 0}
\ee in the leading order. This relation does not depend on which geometric background we are working on.

The idea is that, $\Bbb{S}^d_q$ can be conformally mapped to $\Bbb{H}^1\times\Bbb{S}^{d-1}_q$, therefore the R\'enyi entropy (or supersymmetric) is invariant~\cite{Casini:2011kv}. In the case with supersymmetry, one has to make sure that in the limit $q\to 0$, the background field on $\Bbb{S}^d_q$ coincides with that on $\Bbb{H}^1\times\Bbb{S}^{d-1}_q$. If that is the case, the asymptotic supersymmetric R\'enyi entropy $S_{q\to 0}$ on $\Bbb{S}^d_q$ will coincide with the minus free energy on $\Bbb{H}^1\times\Bbb{S}^{d-1}_{q\to 0}$. The latter is determined by the supersymmetric Casimir energy~\cite{Assel:2015nca}. We will illustrate the details in the following.

\subsection{From $\Bbb{S}^d_q$ to $\Bbb{H}^{d-p}\times\Bbb{S}^p_q$}
We start with the conformal transformation from conic sphere $\Bbb{S}^d_q$ to hyperbolic space $\Bbb{H}^{d-p}\times\Bbb{S}^p_q$. Of course $\Bbb{S}^d_q$ can be considered as the special case of $p=d$. 

In the particular case $p=1$, the transformation is nothing but the Weyl transformation discussed in~\cite{Casini:2011kv}, which offers a convenient way to compute R\'enyi entropy of CFTs. In this case, the branched $d$-sphere is described as \footnote{We normalize the radius as unit.}
\be\label{brsphere1}
\rmd s^2 = \sin^2\theta q^2\rmd\tau^2 + \rmd\theta^2 + \cos^2\theta \rmd^2\Omega_{d-2}\ ,
\ee with domains of coordinates given by
\be
\tau\in [0,2\pi)\ ,\quad \theta\in \left[0,{\pi\over 2}\right]\ ,
\ee and $\Omega_{d-2}$ is a standard $d$-$2$-dimensional round sphere. The metric (\ref{brsphere1}) can be written as
\be
\rmd s^2 = \sin^2\theta \left(q^2\rmd\tau^2 + {1\over \sin^2\theta}\rmd\theta^2 + \cot^2\theta \rmd^2\Omega_{d-2}\right)\ ,
\ee which can be related to the following space by dropping an overall factor $\sin^2\theta$ and using a coordinate transformation $\cot\theta=\sinh\eta$
\be
\rmd s^2 = q^2\rmd\tau^2 + \rmd\eta^2 + \sinh^2\eta \rmd^2\Omega_{d-2}\ ,
\ee where $\eta\in [0,+\infty)$. This is the space of $\Bbb{H}^{d-1}\times\Bbb{S}^1_q$, which indeed fits the case of $p=1$. 

Now we consider the general cases, $1\leq p< d$. The key observation is that, the branched sphere can be presented in different forms. For instance, we can represent $\Bbb{S}^d_q$ as
\be\label{brsphere2}
\rmd s^2 = \sin^2\theta (\rmd\chi^2+\sin^2\chi q^2\rmd\tau^2) + \rmd\theta^2 + \cos^2\theta \rmd^2\Omega_{d-3}\ ,
\ee with domains
\be
\chi\in [0,\pi]\ ,\quad\tau\in [0,2\pi)\ ,\quad \theta\in \left[0,{\pi\over 2}\right]\ ,
\ee and $\Omega_{d-3}$ is a standard $d$-$3$-dimensional round sphere. Again by dropping an overall factor $\sin^2\theta$ and using a coordinate transformation $\cot\theta=\sinh\eta$ for the metric (\ref{brsphere2}), one obtains
\be
\rmd s^2 = \rmd\chi^2+\sin^2\chi q^2\rmd\tau^2 + \rmd\eta^2 + \sinh^2\eta \rmd^2\Omega_{d-3}\ ,
\ee which is the space $\Bbb{H}^{d-2}\times\Bbb{S}^2_q$ with $p=2$. One can follow the same way to eventually figure out the Weyl transformations between $\Bbb{S}^d_q$ and $\Bbb{H}^{d-p}\times\Bbb{S}^p_q$ for any integer $1\leq p< d$.

Since the R\'enyi entropy on $\Bbb{S}^d_q$ can not depend on which particular circle we choose to create the conical singularity, one eventually arrives at the conclusion by employing the same argument in~\cite{Casini:2011kv}:~\footnote{Again by the universal part of R\'enyi entropy we refer to the scheme independent part.} 

{\it The universal part of CFT$_d$ R\'enyi entropy is invariant on $\Bbb{H}^{d-p}\times\Bbb{S}^p_q$ for different integer $p$, where $1\leq p\leq d$.}

For later purpose, let us discuss the particular case $p=d-1$. In this case we describe the branched sphere $\Bbb{S}^d_q$ as
\be\label{brsphere3}
\rmd s^2 = \sin^2\theta (\rmd\chi^2+\sin^2\chi q^2\rmd\tau^2+\cos^2\chi \rmd^2\Omega_{d-3}) + \rmd\theta^2\ ,
\ee with domains
\be
\chi\in \left[0,{\pi\over 2}\right]\ ,\quad\tau\in [0,2\pi)\ ,\quad \theta\in \left[0,\pi\right]\ .
\ee
Again by dropping an overall factor $\sin^2\theta$ for the metric (\ref{brsphere3}), one obtains
\be
\rmd s^2 = \rmd\chi^2+\sin^2\chi q^2\rmd\tau^2+\cos^2\chi \rmd^2\Omega_{d-3} + \rmd\eta^2\ ,
\ee where $\cot\theta=\sinh\eta$ and $\eta\in (-\infty, +\infty)$. This is the space $\Bbb{S}^{d-1}_q\times\Bbb{H}^1$. Here we use $\Bbb{H}^1$ instead of $\Bbb{R}^1$ to emphasize that the volume of $\Bbb{H}^d$ may be regularized.
For free fields, one can compute the CFT R\'enyi entropy on $\Bbb{S}^{d-1}_q\times\Bbb{H}^1$ and show explicitly that the result agrees with that computed from $\Bbb{S}^d_q$ or $\Bbb{S}^1_q\times\Bbb{H}^{d-1}$. In consideration of supersymmetry, one has to add a background field $A_\tau$ along the replica $\tau$ circle inside $\Bbb{S}^{d-1}_q$, in order to find the agreement.

\subsection{Coincidence of backgrounds}
Our main concern is physical quantities for CFTs. For this purpose we can work on $\Bbb{S}^{d-1}_{\sqrt{q}}\times\Bbb{H}^1_{1/\sqrt{q}}$ instead of $\Bbb{S}^{d-1}_q\times\Bbb{H}^1$ because they are related by a scale transformation
\be
{1\over \sqrt{q}} [\Bbb{S}^{d-1}_q\times\Bbb{H}^1]\, =\,  [\Bbb{S}^{d-1}_{\sqrt{q}}\times\Bbb{H}^1_{1/\sqrt{q}}]\ .
\ee
Furthermore, we focus on the limit $q\to 0$. For this purpose, one can instead consider $\Bbb{S}^{d-1}_{\sqrt{q}}\times\Bbb{S}^1_{1/\sqrt{q}}$ because it is equivalent to $\Bbb{S}^{d-1}_{\sqrt{q}}\times\Bbb{H}^1_{1/\sqrt{q}}$ in the limit $q\to 0$
\be\label{Gequal}
\Bbb{S}^{d-1}_{\sqrt{q}}\times\Bbb{H}^1_{1/\sqrt{q}}\bigg |_{q\to 0} = \Bbb{S}^{d-1}_{\sqrt{q}}\times\Bbb{S}^1_{1/\sqrt{q}}\bigg |_{q\to 0}\ .
\ee
In consideration of supersymmetry, one can use the squashed sphere $\widetilde{\Bbb{S}}^{d-1}_{\sqrt{q}}$ to replace the conic sphere $\Bbb{S}^{d-1}_{\sqrt{q}}$ in the right hand side of (\ref{Gequal}), because supersymmetric partition functions do not depend on the resolving factor~\cite{Huang:2014gca, Hama:2014iea,Alday:2013lba,Closset:2013vra,Closset:2014uda,Assel:2014paa
}.\footnote{For this reason, we will not distinguish $d$-$1$-dimensional squashed sphere and conic sphere in the following unless it is necessary.} (\ref{Gequal}) is useful in the sense that it offers a way to compute the asymptotic supersymmetric R\'enyi entropy for interacting SCFTs. To do this, one has to make sure that the background gauge field on $\Bbb{S}^{d-1}_{\sqrt{q}}\times\Bbb{S}^1_{1/\sqrt{q}}$ agrees with that on the original space $\Bbb{S}^d_q$. Fortunately we have more knowledge about supersymmetric partition functions on $\Bbb{S}^{d-1}\times\Bbb{S}^1$ or its generalized version $\Bbb{S}^{d-1}_b\times\Bbb{S}^1_\beta$, where $b$ is the squashing parameter.

\subsection{Squashed Casimir energy}
Now we make a connection between the asymptotic R\'enyi entropy and Casimir energy. It is known that the partition function $Z$ on $\Bbb{S}^{d-1}_b\times\Bbb{S}^1_\beta$ is determined by the Casimir energy on $\Bbb{S}^{d-1}_b$ in the limit $\beta\to \infty$
\be
E_c := -\lim_{\beta\to\infty}\partial_\beta \log Z(\beta)\ ,
\ee which is equivalent to say
\be
\lim_{\beta\to \infty}\log Z(\beta) = -\beta E_c\ .
\ee
In this work, we concern the case with supersymmetry. In the particular case of $6d$ $(2,0)$ theories, the supersymmetric Casimir energy has been studied in~\cite{Bobev:2015kza} \footnote{For the $6d$ $(2,0)$ superconformal index, see~\cite{Bhattacharya:2008zy,Kim:2012qf,Lockhart:2012vp}.}, where the authors considered a general 5-sphere with squashing parameters $\vec\omega = (\omega_1,\omega_2,\omega_3)$. The squashing parameters are defined as parameters appearing in the Killing vector
\be
K = \omega_1{\partial\over\partial\phi_1} + \omega_2{\partial\over\partial\phi_2} + \omega_3{\partial\over\partial\phi_3}\ ,
\ee where $\phi_1,\phi_2, \phi_3$ are three circles representing $U(1)^3$ isometries of $\Bbb{S}^5$. The supersymmetric Casimir energy of an interacting $(2,0)$ theory is~\cite{Bobev:2015kza}
\be\label{casimirInteracting}
E_\mathfrak{g} = r_\mathfrak{g} E_{\mathfrak{u}(1)}- d_\mathfrak{g} h^\vee_\mathfrak{g}{\sigma_1^2\sigma_2^2\over 24\,\omega_1\omega_2\omega_3}\ ,
\ee where $\sigma_1$ and $\sigma_2$ are chemical potentials for the two Cartans of the $SO(5)$ R-symmetry and $E_{\mathfrak{u}(1)}$ is given by
\be
E_{\mathfrak{u}(1)}=-{1\over 48\omega_1\omega_2\omega_3}\left[\sigma_1^2\sigma_2^2-\sum_{i<j}\omega_i^2\omega_j^2 + {1\over 4}\left(\sum_j \omega_j^2-\sigma_1^2-\sigma_2^2\right)^2\right]\ .
\ee 
For the particular case of $\Bbb{S}^5_q\times\Bbb{S}^1$ (which is equivalent to $\Bbb{S}^5_{\sqrt{q}}\times\Bbb{S}^1_{1\over \sqrt{q}}$ for CFTs), we should identify the shape parameters as
\be
\omega_1=\omega_2=1\ ,\quad \omega_3={1\over q}\ .
\ee
In the limit $q\to 0$, in order to match our chemical potentials (\ref{chemicalback}), we set $\sigma_1$ and $\sigma_2$ as \footnote{The $q$ scalings in chemical potentials appear following the convention in~\cite{Bobev:2015kza}.}
\be
\sigma^2_1(q\to 0) = {r_1^2\over q^2}\ ,\quad \sigma^2_2(q\to 0) = {r_2^2\over q^2}\ ,\quad \text{with}~~r_1+r_2 =1\ .
\ee
Evaluating (\ref{casimirInteracting}) we get
\be
E_\mathfrak{g}\bigg |_{q\to 0} = -{1\over 24}{r_1^2r_2^2\over q^3}(r_\mathfrak{g}+d_\mathfrak{g} h^\vee_\mathfrak{g})\ .
\ee
Therefore the free energy \footnote{$f:={I\over V}$.}
\be
f[\Bbb{S}^5_{q\to 0}\times\Bbb{S}^1]={1\over \pi^3}\beta E_\mathfrak{g}\bigg |_{q\to 0} =- {1\over 12\pi^2}{r_1^2r_2^2\over q^3}(r_\mathfrak{g}+d_\mathfrak{g} h^\vee_\mathfrak{g})\ ,
\ee where we have divided a $q$-independent volume factor Vol $[\Bbb{D}^4\times\Bbb{S}^1]=\pi^3$.
Due to (\ref{Gequal}), we have
\be\label{fconnection}
f[\Bbb{S}^5_{q\to 0}\times\Bbb{S}^1]=f[\Bbb{S}^1_{q\to 0}\times\Bbb{H}^5]\ ,
\ee from which we obtain the asymptotic supersymmetric R\'enyi entropy on $\Bbb{S}^1_q\times\Bbb{H}^5$
\be\label{renyiasympt}
S_{q\to 0} = -I_{q\to 0} = {1\over 12}{r_1^2r_2^2\over q^3}(r_\mathfrak{g}+d_\mathfrak{g} h^\vee_\mathfrak{g})\ .
\ee
This fixes the undetermined coefficient $A$ in (\ref{ansatz}) as
\be
A = r_\mathfrak{g}+ h^\vee_\mathfrak{g}d_\mathfrak{g}\ .
\ee
Notice that the fact that the free limit of (\ref{renyiasympt}) precisely agrees with the leading large $\gamma$ term of (\ref{freeformula}) by itself is nontrivial, which confirms the validity of (\ref{fconnection}) in the free case.

\section{Large $N$ limit}
\label{SREholography}
In the large $N$ limit of the $(2,0)$ theory with $\mathfrak{g}=A_{N-1}$, the supersymmetric R\'enyi entropy (\ref{SREfull}) becomes
\ba
{S^{(2,0)}_\gamma\over N^3}&=&\frac{1}{12} r_1^2r_2^2 \,(\gamma -1)^3 +\frac{4}{12} r_1r_2\,(\gamma -1)^2\nn
&+&\frac{4}{12} (1+2r_1r_2) \,(\gamma -1)+\frac{4}{3}\ .\label{largeNSRE}
\ea
We will demonstrate in this section that the above large $N$ result precisely agrees with the holographic result from the seven-dimensional BPS topological black hole in gauged supergravity.

\subsection{Gauged supergravity}
The seven-dimensional gauged $SO(5)$ supergravity can be obtained by Kaluza-Klein reduction of eleven-dimensional supergravity on $\Bbb{S}^4$. For our purpose, we consider a truncation where only the metic, two gauge fields associated to two Cartans of $SO(5)$ and two scalars are retained. The seven-dimensional Lagrangian is given by~\cite{Cvetic:1999xp}
\be\label{lang}
{1\over \sqrt{g}} {\cal L} = R - {1\over 2}(\partial\vec\phi)^2 - {4\over L^2}V - {1\over 4}\sum_{i=1}^2 {1\over X_i^2}\left(F^i_{(2)}\right)^2\ ,
\ee where $\vec\phi=(\phi_1,\phi_2)$ are two scalars and 
\be
X_i = e^{-{1\over 2}\vec a_i\cdot \vec\phi}\ ,i=1,2\ . \quad \vec a_1 = \left(\sqrt{2}, \sqrt{2\over 5}\right)\ ,\quad \vec a_2 = \left(-\sqrt{2}, \sqrt{2\over 5}\right)\ .
\ee The potential is given by
\be
V = -4X_1X_2-2X_0X_1-2X_0X_2+ {1\over 2}X_0^2\ ,\quad X_0 = {1\over X_1X_2}\ .
\ee
Note that for two equal scalars and two equal gauge strengths, the Lagrangian (\ref{lang}) can be further truncated. Turn to the CFT side,
6d $(2,0)$ theories have global $SO(5)$ R-symmetry, which corresponds to the $SO(5)$ gauge group in the bulk supergravity. Also there could be two $U(1)$ background fields used to compensate the singularity on $\Bbb{S}_q^6$, which correspond to $A^1, A^2$ in the gauged supergravity.

\subsection{Topological black hole}
The $2$-charge $7d$ AdS black hole solution for (\ref{lang}) was found in~\cite{Cvetic:1999xp}
\ba\label{tbhseven}
\rmd s_7^2 &=& -{1\over [h_1h_2]^{4\over 5}}\,f(r) \rmd t^2 +[h_1h_2]^{1\over 5} \left({\rmd r^2\over f(r)} + r^2 \rmd\Omega_{5,k}^2\right) \nn & &f(r)=k-{m\over r^4}+{r^2\over L^2}h_1h_2\ , \quad h_i=1+{q_i\over r^4}\ ,
\ea
together with scalars and gauge fields
\be\label{scalargaugeseven}
X_i = {[h_1h_2]^{2\over 5}\over h_i}\ ,\quad A^i = \left[\sqrt{k}\left({1\over h_i}-1\right)+\mu_i\right]\rmd t\ .
\ee
$\rmd\Omega^2_{5,k}$ is the metric on a unit $\Bbb{S}^5$, $\Bbb{T}^5$ or $\Bbb{H}^5$ corresponding to $k=1,0,-1$, respectively.  Since our concern is the $6d$ SCFT on $\Bbb{S}^1\times\Bbb{H}^5$, we are particularly interested in the extremal solution with hyperbolic foliation, where $m=0$ and $k=-1$. We will first proceed in Lorentz signature and assume a well-defined Wick rotation. 

The solution (\ref{tbhseven}) is a BPS topological black hole with two charges. For convenience, define a rescaled charge
\be
\kappa_i = {q_i\over r^4_{H}}\ ,
\ee where the horizon $r_H$ is the largest root of the equation 
\be
f(r_H) = 0\ .
\ee
Then the horizon can be expressed in terms of $\kappa_i$
\be\label{horseven}
r_H = {L\over \sqrt{(1+\kappa_1)(1+\kappa_2)}}\ .
\ee
The Hawking temperature of this black hole is
\ba\label{Tseven}
T &=& {f'(r)\over 4\pi \sqrt{h_1h_2}}\bigg |_{r=r_{H}} \nn &=& {1-\kappa_1-\kappa_2-3\kappa_1\kappa_2\over 2\pi L(1+\kappa_1)(1+\kappa_2)}\ .
\ea
When all charges vanish, we get to the temperature of the uncharged black hole
\be\label{Tzeroseven}
T_0 = {1\over 2\pi L}\ .
\ee
The Bekenstein-Hawking entropy is given by the outer horizon area
\be\label{Sseven}
S = {V_5L^5\over 4G_7}{1\over (1+\kappa_1)^2(1+\kappa_2)^2}\ ,
\ee
where $G_7$ is the seven dimensional Newton constant and $V_5$ is the regularized volume of $\Bbb{H}^5$. The total charge $Q_i$ can be computed by Gauss law
\ba\label{Qiseven}
Q_i &=& {1\over 16\pi G_7} \int_{r\to \infty} -\sqrt{g}F^{rt} = {V_5\over 4\pi G_7} i q_i \nn &=& {V_5L^4\over 4\pi G_7}{i\kappa_i\over (1+\kappa_1)^2(1+\kappa_2)^2}\ .
\ea
The chemical potential is
\be\label{muiseven}
\mu_i = {i\over \kappa_i^{-1}+1}\ .
\ee

\subsection{Precise check}
To match the background gauge fields of the boundary CFT, we set
\be
\mu_1=i(1-\gamma){r_1\over 2}\ ,\quad \mu_2 = i(1-\gamma){r_2\over 2}\ , \quad \text{with}\quad r_1+r_2=1\ .
\ee
By using these inputs, we can solve $\kappa_1$ and $\kappa_2$ by (\ref{muiseven}). Then all physical quantities $T, S, Q_i$ can be worked out explicitly.
One can eventually compute the holographic supersymmetric R\'enyi entropy using the formula derived in~\cite{Huang:2014gca}
\be
S_q = {q\over 1-q}\int _q^1 \left({S(n)\over n^2} - {Q_i(n)\mu_i^\prime(n)\over T_0}\right) \rmd n\ .
\ee Written in terms of $\gamma:=1/q$, the result is given by
\be\label{hsre}
S_{\gamma} = \frac{L^5 V_5}{4 G_7}\left[\frac{r_1^2r_2^2 (\gamma -1)^3}{16}+\frac{(1+2r_1r_2) (\gamma -1)}{4}+\frac{(\gamma -1)^2 r_1r_2}{4}+1\right]\ .
\ee
By identifying the bulk and boundary parameters,
\be
\frac{L^5 V_5}{4 G_7}={4\over 3}N^3\ ,
\ee
one can write the holographic result as
\be
S_{\gamma} =N^3\left( \frac{r_1^2r_2^2 (\gamma -1)^3}{12}+\frac{(1+2r_1r_2) (\gamma -1)}{3}+\frac{(\gamma -1)^2 r_1r_2}{3}+\frac{4}{3}\right)\ .
\ee This precisely agrees with the field theory result (\ref{largeNSRE}).

\section{A possible $a/c$ bound}
As what has been observed in $4d$ SCFTs~\cite{Zhou:2015cpa}, the R\'enyi entropy inequalities indicate the $a/c$ bounds in field theories \footnote{The validity of these inequalities for supersymmetric R\'enyi entropy is expected although a proof is still in preparation.},
\ba
\partial_q H_q \leq 0\ ,\label{ineq1}\\
\partial_q\left({q-1\over q}H_q\right)\geq 0\ ,\label{ineq2}\\
\partial_q((q-1)H_q)\geq 0\ ,\label{ineq3}\\
\partial_q^2((q-1)H_q)\leq 0\label{ineq4}\ ,
\ea where $H_q:=S_q/S_1$.
Imposing these conditions to our results (\ref{fullSRE})(\ref{boundtest0})(\ref{boundtest}), one obtains
\be\label{bounds}
0< {\bar c\over \bar a}\leq {7\over 3}\ ,
\ee or equivalently
\be\label{bounds1}
{\bar a\over \bar c}\geq {3\over 7}\ .
\ee
Note that all the $a,c$ data of the currently known $6d$ $(2,0)$ SCFTs, listed in Table \ref{acdata} in Appendix \ref{appA}, satisfy the inequality (\ref{bounds})(\ref{bounds1}).
The lowest $\bar a / \bar c$ value in the current data, $4/7$, supported by the large $N$ limits, is greater than our bound $3/7$. Note that the expression of supersymmetric R\'enyi entropy in terms of $a,c$ anomalies could work for theories beyond the ADE type. It would be interesting to understand whether our bound implies new $(2,0)$ SCFTs.
It would also be interesting to understand similar bounds in SCFTs with less supersymmetry. We leave these questions for future work.
\section*{Acknowledgement}
The author is grateful for helpful discussions with Ofer Aharony, Thomas Dumitrescu, Igor Klebanov, Zohar Komargodski, Hong Liu, Mark Mezei, Jun Nian, Eric Perlmutter, Soo Jong Rey, Amit Sever, Cobi Sonnenschein and Xi Yin. The author would like to thank Princeton University and Harvard University for hospitality. This work was supported by ``The PBC program of the Israel council of higher education'' and in part by the Israel Science Foundation (grant 1989/14), the US-Israel bi-national fund (BSF) grant 2012383 and the German Israel bi-national fund GIF grant number I-244-303.7-2013.

\appendix

\section{Data of simply-laced Lie algebra $\mathfrak{g}$}
\label{appA}

\begin{table}[htp]
\caption{The rank $r_\mathfrak{g}$, dual Coxeter number $h^\vee_\mathfrak{g}$, dimension $d_\mathfrak{g}$ of the simply-laced Lie algebras and the normalized $a,c$ anomalies for the associated $6d$ $(2,0)$ SCFTs~\cite{Cordova:2015vwa}.}
\begin{center}
\begin{tabular}{l*{7}{c}r}
$\mathfrak{g}$ & $r_{\mathfrak{g}}$ & $h^\vee_{\mathfrak{g}}$ & $d_{\mathfrak{g}}$ & $\bar a_{\mathfrak{g}}$ & $\bar c_{\mathfrak{g}}$ & $\bar a_{\mathfrak{g}}/ \bar c_{\mathfrak{g}}$\\
\hline
$A_{n-1}$ & $n-1$ & $n$ & $n^2-1$ & ${16\over 7}n^3-{9\over 7}n-1$ & $4n^3-3n-1$ & $\frac{3}{7 (2 n+1)^2}+\frac{4}{7}$\\
$D_{n}$ & $n$ & $2n-2$ & $n(2n-1)$ & ${64\over 7}n^3-{96\over 7}n^2+{39\over 7}n$ & $16n^3-24n^2+9n$ & $\frac{3}{7 (3-4 n)^2}+\frac{4}{7}$\\
$E_6$ & $6$ & $12$ & $78$ & ${15018\over 7}$ & $3750$ & $\sim 0.572114$\\
$E_7$ & $7$ & $18$ & $133$ & $5479$ & $9583$ & $\sim 0.571742$\\
$E_8$ & $8$ & $30$ & $248$ & ${119096\over 7}$ & $29768$ & $\sim 0.571544>\frac{4}{7}$\\
\end{tabular}
\end{center}
\label{acdata}
\end{table}


\begin{thebibliography}{100}

%\cite{Nahm:1977tg}
\bibitem{Nahm:1977tg} 
  W.~Nahm,
  ``Supersymmetries and their Representations,''
  Nucl.\ Phys.\ B {\bf 135}, 149 (1978).
  doi:10.1016/0550-3213(78)90218-3
  %%CITATION = doi:10.1016/0550-3213(78)90218-3;%%
  %523 citations counted in INSPIRE as of 29 Nov 2015
  
\bibitem{Cordova}
C.Cordova,
``Applications of Superconformal Representation Theory,''
http://www.birs.ca/events/2015/5-day-workshops/15w5154/videos/watch/201505261651-Cordova.html

%\cite{Gaiotto:2009we}
\bibitem{Gaiotto:2009we} 
  D.~Gaiotto,
  ``N=2 dualities,''
  JHEP {\bf 1208}, 034 (2012)
  doi:10.1007/JHEP08(2012)034
  [arXiv:0904.2715 [hep-th]].
  %%CITATION = doi:10.1007/JHEP08(2012)034;%%
  %505 citations counted in INSPIRE as of 29 Nov 2015
  
%\cite{Gaiotto:2009hg}
\bibitem{Gaiotto:2009hg} 
  D.~Gaiotto, G.~W.~Moore and A.~Neitzke,
  ``Wall-crossing, Hitchin Systems, and the WKB Approximation,''
  arXiv:0907.3987 [hep-th].
  %%CITATION = ARXIV:0907.3987;%%
  %321 citations counted in INSPIRE as of 29 Nov 2015
  
%\cite{Dimofte:2011ju}
\bibitem{Dimofte:2011ju} 
  T.~Dimofte, D.~Gaiotto and S.~Gukov,
  ``Gauge Theories Labelled by Three-Manifolds,''
  Commun.\ Math.\ Phys.\  {\bf 325}, 367 (2014)
  doi:10.1007/s00220-013-1863-2
  [arXiv:1108.4389 [hep-th]].
  %%CITATION = doi:10.1007/s00220-013-1863-2;%%
  %124 citations counted in INSPIRE as of 29 Nov 2015
  
%\cite{Bah:2012dg}
\bibitem{Bah:2012dg} 
  I.~Bah, C.~Beem, N.~Bobev and B.~Wecht,
  ``Four-Dimensional SCFTs from M5-Branes,''
  JHEP {\bf 1206}, 005 (2012)
  doi:10.1007/JHEP06(2012)005
  [arXiv:1203.0303 [hep-th]].
  %%CITATION = doi:10.1007/JHEP06(2012)005;%%
  %73 citations counted in INSPIRE as of 29 Nov 2015
  
%\cite{Gadde:2013sca}
\bibitem{Gadde:2013sca} 
  A.~Gadde, S.~Gukov and P.~Putrov,
  ``Fivebranes and 4-manifolds,''
  arXiv:1306.4320 [hep-th].
  %%CITATION = ARXIV:1306.4320;%%
  %34 citations counted in INSPIRE as of 29 Nov 2015
  
  %\cite{Lambert:2010wm}
\bibitem{Lambert:2010wm} 
  N.~Lambert and C.~Papageorgakis,
  ``Nonabelian (2,0) Tensor Multiplets and 3-algebras,''
  JHEP {\bf 1008}, 083 (2010)
  doi:10.1007/JHEP08(2010)083
  [arXiv:1007.2982 [hep-th]].
  %%CITATION = doi:10.1007/JHEP08(2010)083;%%
  %72 citations counted in INSPIRE as of 06 Dec 2015
  
  %\cite{Ho:2011ni}
\bibitem{Ho:2011ni} 
  P.~M.~Ho, K.~W.~Huang and Y.~Matsuo,
  ``A Non-Abelian Self-Dual Gauge Theory in 5+1 Dimensions,''
  JHEP {\bf 1107}, 021 (2011)
  doi:10.1007/JHEP07(2011)021
  [arXiv:1104.4040 [hep-th]].
  %%CITATION = doi:10.1007/JHEP07(2011)021;%%
  %52 citations counted in INSPIRE as of 06 Dec 2015
  
 %\cite{Chu:2012um}
\bibitem{Chu:2012um} 
  C.~S.~Chu and S.~L.~Ko,
  ``Non-abelian Action for Multiple Five-Branes with Self-Dual Tensors,''
  JHEP {\bf 1205}, 028 (2012)
  doi:10.1007/JHEP05(2012)028
  [arXiv:1203.4224 [hep-th]].
  %%CITATION = doi:10.1007/JHEP05(2012)028;%%
  %41 citations counted in INSPIRE as of 06 Dec 2015
  
  %\cite{Bonetti:2012st}
\bibitem{Bonetti:2012st} 
  F.~Bonetti, T.~W.~Grimm and S.~Hohenegger,
  ``Non-Abelian Tensor Towers and (2,0) Superconformal Theories,''
  JHEP {\bf 1305}, 129 (2013)
  doi:10.1007/JHEP05(2013)129
  [arXiv:1209.3017 [hep-th]].
  %%CITATION = doi:10.1007/JHEP05(2013)129;%%
  %29 citations counted in INSPIRE as of 06 Dec 2015
  
 %\cite{Samtleben:2012fb}
\bibitem{Samtleben:2012fb} 
  H.~Samtleben, E.~Sezgin and R.~Wimmer,
  ``Six-dimensional superconformal couplings of non-abelian tensor and hypermultiplets,''
  JHEP {\bf 1303}, 068 (2013)
  doi:10.1007/JHEP03(2013)068
  [arXiv:1212.5199 [hep-th]].
  %%CITATION = doi:10.1007/JHEP03(2013)068;%%
  %20 citations counted in INSPIRE as of 06 Dec 2015
  
  %\cite{Aharony:1997th}
\bibitem{Aharony:1997th} 
  O.~Aharony, M.~Berkooz, S.~Kachru, N.~Seiberg and E.~Silverstein,
  ``Matrix description of interacting theories in six-dimensions,''
  Adv.\ Theor.\ Math.\ Phys.\  {\bf 1}, 148 (1998)
  [hep-th/9707079].
  %%CITATION = HEP-TH/9707079;%%
  %178 citations counted in INSPIRE as of 06 Dec 2015
  
 %\cite{Aharony:1997an}
\bibitem{Aharony:1997an} 
  O.~Aharony, M.~Berkooz and N.~Seiberg,
  ``Light cone description of (2,0) superconformal theories in six-dimensions,''
  Adv.\ Theor.\ Math.\ Phys.\  {\bf 2}, 119 (1998)
  [hep-th/9712117].
  %%CITATION = HEP-TH/9712117;%%
  %171 citations counted in INSPIRE as of 06 Dec 2015
  
 %\cite{ArkaniHamed:2001ie}
\bibitem{ArkaniHamed:2001ie} 
  N.~Arkani-Hamed, A.~G.~Cohen, D.~B.~Kaplan, A.~Karch and L.~Motl,
  ``Deconstructing (2,0) and little string theories,''
  JHEP {\bf 0301}, 083 (2003)
  doi:10.1088/1126-6708/2003/01/083
  [hep-th/0110146].
  %%CITATION = doi:10.1088/1126-6708/2003/01/083;%%
  %119 citations counted in INSPIRE as of 06 Dec 2015
  
  %\cite{Douglas:2010iu}
\bibitem{Douglas:2010iu} 
  M.~R.~Douglas,
  ``On D=5 super Yang-Mills theory and (2,0) theory,''
  JHEP {\bf 1102}, 011 (2011)
  doi:10.1007/JHEP02(2011)011
  [arXiv:1012.2880 [hep-th]].
  %%CITATION = doi:10.1007/JHEP02(2011)011;%%
  %141 citations counted in INSPIRE as of 06 Dec 2015
  
  %\cite{Lambert:2010iw}
\bibitem{Lambert:2010iw} 
  N.~Lambert, C.~Papageorgakis and M.~Schmidt-Sommerfeld,
  ``M5-Branes, D4-Branes and Quantum 5D super-Yang-Mills,''
  JHEP {\bf 1101}, 083 (2011)
  doi:10.1007/JHEP01(2011)083
  [arXiv:1012.2882 [hep-th]].
  %%CITATION = doi:10.1007/JHEP01(2011)083;%%
  %140 citations counted in INSPIRE as of 06 Dec 2015
  
 %\cite{Witten:1988ze}
\bibitem{Witten:1988ze} 
  E.~Witten,
  ``Topological Quantum Field Theory,''
  Commun.\ Math.\ Phys.\  {\bf 117}, 353 (1988).
  doi:10.1007/BF01223371
  %%CITATION = doi:10.1007/BF01223371;%%
  %1288 citations counted in INSPIRE as of 06 Dec 2015
  
 %\cite{Nekrasov:2002qd}
\bibitem{Nekrasov:2002qd} 
  N.~A.~Nekrasov,
  ``Seiberg-Witten prepotential from instanton counting,''
  Adv.\ Theor.\ Math.\ Phys.\  {\bf 7}, no. 5, 831 (2003)
  doi:10.4310/ATMP.2003.v7.n5.a4
  [hep-th/0206161].
  %%CITATION = doi:10.4310/ATMP.2003.v7.n5.a4;%%
  %756 citations counted in INSPIRE as of 06 Dec 2015
  
 %\cite{Pestun:2007rz}
\bibitem{Pestun:2007rz} 
  V.~Pestun,
  ``Localization of gauge theory on a four-sphere and supersymmetric Wilson loops,''
  Commun.\ Math.\ Phys.\  {\bf 313}, 71 (2012)
  doi:10.1007/s00220-012-1485-0
  [arXiv:0712.2824 [hep-th]].
  %%CITATION = doi:10.1007/s00220-012-1485-0;%%
  %572 citations counted in INSPIRE as of 06 Dec 2015
  
  %\cite{Maxfield:2012aw}
\bibitem{Maxfield:2012aw} 
  T.~Maxfield and S.~Sethi,
  ``The Conformal Anomaly of M5-Branes,''
  JHEP {\bf 1206}, 075 (2012)
  doi:10.1007/JHEP06(2012)075
  [arXiv:1204.2002 [hep-th]].
  %%CITATION = doi:10.1007/JHEP06(2012)075;%%
  %26 citations counted in INSPIRE as of 06 Dec 2015
  
%\cite{Cordova:2015vwa}
\bibitem{Cordova:2015vwa} 
  C.~Cordova, T.~T.~Dumitrescu and X.~Yin,
  ``Higher Derivative Terms, Toroidal Compactification, and Weyl Anomalies in Six-Dimensional (2,0) Theories,''
  arXiv:1505.03850 [hep-th].
  %%CITATION = ARXIV:1505.03850;%%
  %11 citations counted in INSPIRE as of 17 Nov 2015
  
%\cite{Beem:2014kka}
\bibitem{Beem:2014kka} 
  C.~Beem, L.~Rastelli and B.~C.~van Rees,
  ``$ \mathcal{W} $ symmetry in six dimensions,''
  JHEP {\bf 1505}, 017 (2015)
  doi:10.1007/JHEP05(2015)017
  [arXiv:1404.1079 [hep-th]].
  %%CITATION = doi:10.1007/JHEP05(2015)017;%%
  %26 citations counted in INSPIRE as of 29 Nov 2015
  
%\cite{Beem:2015aoa}
\bibitem{Beem:2015aoa} 
  C.~Beem, M.~Lemos, L.~Rastelli and B.~C.~van Rees,
  ``The $(2,0)$ superconformal bootstrap,''
  arXiv:1507.05637 [hep-th].
  %%CITATION = ARXIV:1507.05637;%%
  %13 citations counted in INSPIRE as of 29 Nov 2015
  
 %\cite{Deser:1976yx}
\bibitem{Deser:1976yx} 
  S.~Deser, M.~J.~Duff and C.~J.~Isham,
  ``Nonlocal Conformal Anomalies,''
  Nucl.\ Phys.\ B {\bf 111}, 45 (1976).
  doi:10.1016/0550-3213(76)90480-6
  %%CITATION = doi:10.1016/0550-3213(76)90480-6;%%
  %257 citations counted in INSPIRE as of 06 Dec 2015
  
 %\cite{Deser:1993yx}
\bibitem{Deser:1993yx} 
  S.~Deser and A.~Schwimmer,
  ``Geometric classification of conformal anomalies in arbitrary dimensions,''
  Phys.\ Lett.\ B {\bf 309}, 279 (1993)
  doi:10.1016/0370-2693(93)90934-A
  [hep-th/9302047].
  %%CITATION = doi:10.1016/0370-2693(93)90934-A;%%
  %227 citations counted in INSPIRE as of 06 Dec 2015
  
 %\cite{Maldacena:1997re}
\bibitem{Maldacena:1997re} 
  J.~M.~Maldacena,
  ``The Large N limit of superconformal field theories and supergravity,''
  Int.\ J.\ Theor.\ Phys.\  {\bf 38}, 1113 (1999)
  [Adv.\ Theor.\ Math.\ Phys.\  {\bf 2}, 231 (1998)]
  doi:10.1023/A:1026654312961
  [hep-th/9711200].
  %%CITATION = doi:10.1023/A:1026654312961;%%
  %11256 citations counted in INSPIRE as of 05 Dec 2015
  
 %\cite{Witten:1998xy}
\bibitem{Witten:1998xy} 
  E.~Witten,
  ``Baryons and branes in anti-de Sitter space,''
  JHEP {\bf 9807}, 006 (1998)
  [hep-th/9805112].
  %%CITATION = HEP-TH/9805112;%%
  %510 citations counted in INSPIRE as of 06 Dec 2015
  
 %\cite{Aharony:1998rm}
\bibitem{Aharony:1998rm} 
  O.~Aharony, Y.~Oz and Z.~Yin,
  ``M theory on AdS(p) x S(11-p) and superconformal field theories,''
  Phys.\ Lett.\ B {\bf 430}, 87 (1998)
  doi:10.1016/S0370-2693(98)00508-5
  [hep-th/9803051].
  %%CITATION = doi:10.1016/S0370-2693(98)00508-5;%%
  %111 citations counted in INSPIRE as of 06 Dec 2015
  
 %\cite{Henningson:1998gx}
\bibitem{Henningson:1998gx} 
  M.~Henningson and K.~Skenderis,
  ``The Holographic Weyl anomaly,''
  JHEP {\bf 9807}, 023 (1998)
  doi:10.1088/1126-6708/1998/07/023
  [hep-th/9806087].
  %%CITATION = doi:10.1088/1126-6708/1998/07/023;%%
  %1026 citations counted in INSPIRE as of 07 Dec 2015
  
 %\cite{Bastianelli:2000hi}
\bibitem{Bastianelli:2000hi} 
  F.~Bastianelli, S.~Frolov and A.~A.~Tseytlin,
  ``Conformal anomaly of (2,0) tensor multiplet in six-dimensions and AdS / CFT correspondence,''
  JHEP {\bf 0002}, 013 (2000)
  doi:10.1088/1126-6708/2000/02/013
  [hep-th/0001041].
  %%CITATION = doi:10.1088/1126-6708/2000/02/013;%%
  %102 citations counted in INSPIRE as of 05 Dec 2015
  
  %\cite{Bastianelli:1999ab}
\bibitem{Bastianelli:1999ab} 
  F.~Bastianelli, S.~Frolov and A.~A.~Tseytlin,
  ``Three point correlators of stress tensors in maximally supersymmetric conformal theories in D = 3 and D = 6,''
  Nucl.\ Phys.\ B {\bf 578}, 139 (2000)
  doi:10.1016/S0550-3213(99)00822-6
  [hep-th/9911135].
  %%CITATION = doi:10.1016/S0550-3213(99)00822-6;%%
  %34 citations counted in INSPIRE as of 16 Nov 2015
  
 %\cite{Duff:1995wd}
\bibitem{Duff:1995wd} 
  M.~J.~Duff, J.~T.~Liu and R.~Minasian,
  ``Eleven-dimensional origin of string-string duality: A One loop test,''
  Nucl.\ Phys.\ B {\bf 452}, 261 (1995)
  doi:10.1016/0550-3213(95)00368-3
  [hep-th/9506126].
  %%CITATION = doi:10.1016/0550-3213(95)00368-3;%%
  %324 citations counted in INSPIRE as of 06 Dec 2015
  
 %\cite{Witten:1996hc}
\bibitem{Witten:1996hc} 
  E.~Witten,
  ``Five-brane effective action in M theory,''
  J.\ Geom.\ Phys.\  {\bf 22}, 103 (1997)
  doi:10.1016/S0393-0440(97)80160-X
  [hep-th/9610234].
  %%CITATION = doi:10.1016/S0393-0440(97)80160-X;%%
  %379 citations counted in INSPIRE as of 06 Dec 2015
  
  %\cite{Freed:1998tg}
\bibitem{Freed:1998tg} 
  D.~Freed, J.~A.~Harvey, R.~Minasian and G.~W.~Moore,
  ``Gravitational anomaly cancellation for M theory five-branes,''
  Adv.\ Theor.\ Math.\ Phys.\  {\bf 2}, 601 (1998)
  [hep-th/9803205].
  %%CITATION = HEP-TH/9803205;%%
  %104 citations counted in INSPIRE as of 06 Dec 2015
  
%\cite{Harvey:1998bx}
\bibitem{Harvey:1998bx} 
  J.~A.~Harvey, R.~Minasian and G.~W.~Moore,
  ``NonAbelian tensor multiplet anomalies,''
  JHEP {\bf 9809}, 004 (1998)
  doi:10.1088/1126-6708/1998/09/004
  [hep-th/9808060].
  %%CITATION = doi:10.1088/1126-6708/1998/09/004;%%
  %124 citations counted in INSPIRE as of 29 Nov 2015
  
 %\cite{Yi:2001bz}
\bibitem{Yi:2001bz} 
  P.~Yi,
  ``Anomaly of (2,0) theories,''
  Phys.\ Rev.\ D {\bf 64}, 106006 (2001)
  doi:10.1103/PhysRevD.64.106006
  [hep-th/0106165].
  %%CITATION = doi:10.1103/PhysRevD.64.106006;%%
  %35 citations counted in INSPIRE as of 29 Nov 2015
  
%\cite{Intriligator:2000eq}
\bibitem{Intriligator:2000eq} 
  K.~A.~Intriligator,
  ``Anomaly matching and a Hopf-Wess-Zumino term in 6d, N=(2,0) field theories,''
  Nucl.\ Phys.\ B {\bf 581}, 257 (2000)
  doi:10.1016/S0550-3213(00)00148-6
  [hep-th/0001205].
  %%CITATION = doi:10.1016/S0550-3213(00)00148-6;%%
  %68 citations counted in INSPIRE as of 29 Nov 2015

%\cite{Ohmori:2014kda}
\bibitem{Ohmori:2014kda} 
  K.~Ohmori, H.~Shimizu, Y.~Tachikawa and K.~Yonekura,
  ``Anomaly polynomial of general 6d SCFTs,''
  PTEP {\bf 2014}, no. 10, 103B07 (2014)
  doi:10.1093/ptep/ptu140
  [arXiv:1408.5572 [hep-th]].
  %%CITATION = doi:10.1093/ptep/ptu140;%%
  %28 citations counted in INSPIRE as of 29 Nov 2015
  
  %\cite{Casini:2011kv}
\bibitem{Casini:2011kv} 
  H.~Casini, M.~Huerta and R.~C.~Myers,
  ``Towards a derivation of holographic entanglement entropy,''
  JHEP {\bf 1105}, 036 (2011)
  doi:10.1007/JHEP05(2011)036
  [arXiv:1102.0440 [hep-th]].
  %%CITATION = doi:10.1007/JHEP05(2011)036;%%
  %314 citations counted in INSPIRE as of 16 Nov 2015

 %\cite{Nishioka:2013haa}
\bibitem{Nishioka:2013haa} 
  T.~Nishioka and I.~Yaakov,
  ``Supersymmetric Renyi Entropy,''
  JHEP {\bf 1310}, 155 (2013)
  doi:10.1007/JHEP10(2013)155
  [arXiv:1306.2958 [hep-th]].
  %%CITATION = doi:10.1007/JHEP10(2013)155;%%
  %25 citations counted in INSPIRE as of 01 Dec 2015
  
%\cite{Huang:2014gca}
\bibitem{Huang:2014gca} 
  X.~Huang, S.~J.~Rey and Y.~Zhou,
  ``Three-dimensional SCFT on conic space as hologram of charged topological black hole,''
  JHEP {\bf 1403}, 127 (2014)
  doi:10.1007/JHEP03(2014)127
  [arXiv:1401.5421 [hep-th]].
  %%CITATION = doi:10.1007/JHEP03(2014)127;%%
  %15 citations counted in INSPIRE as of 01 Dec 2015
  
 %\cite{Nishioka:2014mwa}
\bibitem{Nishioka:2014mwa} 
  T.~Nishioka,
  ``The Gravity Dual of Supersymmetric Renyi Entropy,''
  JHEP {\bf 1407}, 061 (2014)
  doi:10.1007/JHEP07(2014)061
  [arXiv:1401.6764 [hep-th]].
  %%CITATION = doi:10.1007/JHEP07(2014)061;%%
  %9 citations counted in INSPIRE as of 01 Dec 2015 
  
   %\cite{Huang:2014pda}
  \bibitem{Huang:2014pda} 
  X.~Huang and Y.~Zhou,
  ``$ \mathcal{N}=4 $ Super-Yang-Mills on conic space as hologram of STU topological black hole,''
  JHEP {\bf 1502}, 068 (2015)
  doi:10.1007/JHEP02(2015)068
  [arXiv:1408.3393 [hep-th]].
  %%CITATION = doi:10.1007/JHEP02(2015)068;%%
  %9 citations counted in INSPIRE as of 16 Nov 2015
  
%\cite{Zhou:2015cpa}
\bibitem{Zhou:2015cpa} 
  Y.~Zhou,
  ``Universal Features of Four-Dimensional Superconformal Field Theory on Conic Space,''
  JHEP {\bf 1508}, 052 (2015)
  doi:10.1007/JHEP08(2015)052
  [arXiv:1506.06512 [hep-th]].
  %%CITATION = doi:10.1007/JHEP08(2015)052;%%
  %2 citations counted in INSPIRE as of 01 Dec 2015
  
 %\cite{Hofman:2008ar}
\bibitem{Hofman:2008ar} 
  D.~M.~Hofman and J.~Maldacena,
  ``Conformal collider physics: Energy and charge correlations,''
  JHEP {\bf 0805}, 012 (2008)
  doi:10.1088/1126-6708/2008/05/012
  [arXiv:0803.1467 [hep-th]].
  %%CITATION = doi:10.1088/1126-6708/2008/05/012;%%
  %244 citations counted in INSPIRE as of 06 Dec 2015 
  
%cite{Crossley:2014oea}
\bibitem{Crossley:2014oea} 
  M.~Crossley, E.~Dyer and J.~Sonner,
  ``Super-Renyi entropy, Wilson loops for N=4 SYM and their gravity duals,''
  JHEP {\bf 1412}, 001 (2014)
  doi:10.1007/JHEP12(2014)001
  [arXiv:1409.0542 [hep-th]].
  %%CITATION = doi:10.1007/JHEP12(2014)001;%%
  %10 citations counted in INSPIRE as of 01 Dec 2015
  
%\cite{Alday:2014fsa}
\bibitem{Alday:2014fsa} 
  L.~F.~Alday, P.~Richmond and J.~Sparks,
  ``The holographic supersymmetric Renyi entropy in five dimensions,''
  JHEP {\bf 1502}, 102 (2015)
  doi:10.1007/JHEP02(2015)102
  [arXiv:1410.0899 [hep-th]].
  %%CITATION = doi:10.1007/JHEP02(2015)102;%%
  %7 citations counted in INSPIRE as of 01 Dec 2015
  
  %\cite{Hama:2014iea}
\bibitem{Hama:2014iea} 
  N.~Hama, T.~Nishioka and T.~Ugajin,
  ``Supersymmetric Renyi entropy in five dimensions,''
  JHEP {\bf 1412}, 048 (2014)
  doi:10.1007/JHEP12(2014)048
  [arXiv:1410.2206 [hep-th]].
  %%CITATION = doi:10.1007/JHEP12(2014)048;%%
  %6 citations counted in INSPIRE as of 01 Dec 2015
  
%\cite{Nian:2015xky}
\bibitem{Nian:2015xky} 
  J.~Nian and Y.~Zhou,
  ``Renyi Entropy of Free (2,0) Tensor Multiplet and its Supersymmetric Counterpart,''
  arXiv:1511.00313 [hep-th].
  %%CITATION = ARXIV:1511.00313;%%
  
 %\cite{Giveon:2015cgs}
\bibitem{Giveon:2015cgs} 
  A.~Giveon and D.~Kutasov,
  ``Supersymmetric Renyi Entropy in $CFT_2$ and $AdS_3$,''
  arXiv:1510.08872 [hep-th].
  %%CITATION = ARXIV:1510.08872;%%
  
%\cite{Perlmutter:2013gua}
\bibitem{Perlmutter:2013gua} 
  E.~Perlmutter,
  ``A universal feature of CFT Renyi entropy,''
  JHEP {\bf 1403}, 117 (2014)
  doi:10.1007/JHEP03(2014)117
  [arXiv:1308.1083 [hep-th]].
  %%CITATION = doi:10.1007/JHEP03(2014)117;%%
  %24 citations counted in INSPIRE as of 16 Nov 2015
  
 %\cite{Lee:2014zaa}
\bibitem{Lee:2014zaa} 
  J.~Lee, A.~Lewkowycz, E.~Perlmutter and B.~R.~Safdi,
  ``Renyi entropy, stationarity, and entanglement of the conformal scalar,''
  JHEP {\bf 1503}, 075 (2015)
  doi:10.1007/JHEP03(2015)075
  [arXiv:1407.7816 [hep-th]].
  %%CITATION = doi:10.1007/JHEP03(2015)075;%%
  %15 citations counted in INSPIRE as of 16 Nov 2015
  
 %\cite{Hung:2014npa}
\bibitem{Hung:2014npa} 
  L.~Y.~Hung, R.~C.~Myers and M.~Smolkin,
  ``Twist operators in higher dimensions,''
  JHEP {\bf 1410}, 178 (2014)
  doi:10.1007/JHEP10(2014)178
  [arXiv:1407.6429 [hep-th]].
  %%CITATION = doi:10.1007/JHEP10(2014)178;%%
  %26 citations counted in INSPIRE as of 06 Dec 2015
  
  %\cite{Belin:2013uta}
\bibitem{Belin:2013uta} 
  A.~Belin, L.~Y.~Hung, A.~Maloney, S.~Matsuura, R.~C.~Myers and T.~Sierens,
  ``Holographic Charged Renyi Entropies,''
  JHEP {\bf 1312}, 059 (2013)
  doi:10.1007/JHEP12(2013)059
  [arXiv:1310.4180 [hep-th]].
  %%CITATION = doi:10.1007/JHEP12(2013)059;%%
  %19 citations counted in INSPIRE as of 16 Nov 2015
  
 %\cite{Bergshoeff:1999db}
\bibitem{Bergshoeff:1999db} 
  E.~Bergshoeff, E.~Sezgin and A.~Van Proeyen,
  ``(2,0) tensor multiplets and conformal supergravity in D = 6,''
  Class.\ Quant.\ Grav.\  {\bf 16}, 3193 (1999)
  doi:10.1088/0264-9381/16/10/311
  [hep-th/9904085].
  %%CITATION = doi:10.1088/0264-9381/16/10/311;%%
  %29 citations counted in INSPIRE as of 05 Dec 2015
  
 %\cite{Osborn:1993cr}
\bibitem{Osborn:1993cr} 
  H.~Osborn and A.~C.~Petkou,
  ``Implications of conformal invariance in field theories for general dimensions,''
  Annals Phys.\  {\bf 231}, 311 (1994)
  doi:10.1006/aphy.1994.1045
  [hep-th/9307010].
  %%CITATION = doi:10.1006/aphy.1994.1045;%%
  %296 citations counted in INSPIRE as of 02 Dec 2015
  
 %\cite{Erdmenger:1996yc}
\bibitem{Erdmenger:1996yc} 
  J.~Erdmenger and H.~Osborn,
  ``Conserved currents and the energy momentum tensor in conformally invariant theories for general dimensions,''
  Nucl.\ Phys.\ B {\bf 483}, 431 (1997)
  doi:10.1016/S0550-3213(96)00545-7
  [hep-th/9605009].
  %%CITATION = doi:10.1016/S0550-3213(96)00545-7;%%
  %160 citations counted in INSPIRE as of 06 Dec 2015
  
%\cite{Eden:2001wg}
\bibitem{Eden:2001wg} 
  B.~Eden, S.~Ferrara and E.~Sokatchev,
  ``(2,0) superconformal OPEs in D = 6, selection rules and nonrenormalization theorems,''
  JHEP {\bf 0111}, 020 (2001)
  doi:10.1088/1126-6708/2001/11/020
  [hep-th/0107084].
  %%CITATION = doi:10.1088/1126-6708/2001/11/020;%%
  %20 citations counted in INSPIRE as of 05 Dec 2015
  
 %\cite{Arutyunov:2002ff}
\bibitem{Arutyunov:2002ff} 
  G.~Arutyunov and E.~Sokatchev,
  ``Implications of superconformal symmetry for interacting (2,0) tensor multiplets,''
  Nucl.\ Phys.\ B {\bf 635}, 3 (2002)
  doi:10.1016/S0550-3213(02)00359-0
  [hep-th/0201145].
  %%CITATION = doi:10.1016/S0550-3213(02)00359-0;%%
  %20 citations counted in INSPIRE as of 05 Dec 2015
  
 %\cite{Dolan:2004mu}
\bibitem{Dolan:2004mu} 
  F.~A.~Dolan, L.~Gallot and E.~Sokatchev,
  ``On four-point functions of 1/2-BPS operators in general dimensions,''
  JHEP {\bf 0409}, 056 (2004)
  doi:10.1088/1126-6708/2004/09/056
  [hep-th/0405180].
  %%CITATION = doi:10.1088/1126-6708/2004/09/056;%%
  %33 citations counted in INSPIRE as of 05 Dec 2015
  
%\cite{Witten:1995zh}
\bibitem{Witten:1995zh} 
  E.~Witten,
  ``Some comments on string dynamics,''
  In *Los Angeles 1995, Future perspectives in string theory* 501-523
  [hep-th/9507121].
  %%CITATION = HEP-TH/9507121;%%
  %518 citations counted in INSPIRE as of 29 Nov 2015
  
%\cite{Strominger:1995ac}
\bibitem{Strominger:1995ac} 
  A.~Strominger,
  ``Open p-branes,''
  Phys.\ Lett.\ B {\bf 383}, 44 (1996)
  doi:10.1016/0370-2693(96)00712-5
  [hep-th/9512059].
  %%CITATION = doi:10.1016/0370-2693(96)00712-5;%%
  %528 citations counted in INSPIRE as of 29 Nov 2015
  
 %\cite{Witten:1995em}
\bibitem{Witten:1995em} 
  E.~Witten,
  ``Five-branes and M theory on an orbifold,''
  Nucl.\ Phys.\ B {\bf 463}, 383 (1996)
  doi:10.1016/0550-3213(96)00032-6
  [hep-th/9512219].
  %%CITATION = doi:10.1016/0550-3213(96)00032-6;%%
  %240 citations counted in INSPIRE as of 29 Nov 2015
  
 \bibitem{MooreLecute} 
 G.~W.~Moore,
 ``Lecute Notes for Felix Klein Lectures,''
 http://www.physics.rutgers.edu/~gmoore/FelixKleinLectureNotes.pdf
 
 %\cite{Aharony:2015zea}
\bibitem{Aharony:2015zea} 
  O.~Aharony, M.~Berkooz and S.~J.~Rey,
  ``Rigid holography and six-dimensional $ \mathcal{N}=\left(2,0\right) $ theories on AdS$_{5}\times \mathbb{S}^{1}$,''
  JHEP {\bf 1503}, 121 (2015)
  doi:10.1007/JHEP03(2015)121
  [arXiv:1501.02904 [hep-th]].
  %%CITATION = doi:10.1007/JHEP03(2015)121;%%
  %7 citations counted in INSPIRE as of 04 Dec 2015 
  
 %\cite{Casini:2010kt}
\bibitem{Casini:2010kt} 
  H.~Casini and M.~Huerta,
  ``Entanglement entropy for the n-sphere,''
  Phys.\ Lett.\ B {\bf 694}, 167 (2011)
  doi:10.1016/j.physletb.2010.09.054
  [arXiv:1007.1813 [hep-th]].
  %%CITATION = doi:10.1016/j.physletb.2010.09.054;%%
  %58 citations counted in INSPIRE as of 06 Dec 2015 
  
 %\cite{Klebanov:2011uf}
\bibitem{Klebanov:2011uf} 
  I.~R.~Klebanov, S.~S.~Pufu, S.~Sachdev and B.~R.~Safdi,
  ``Renyi Entropies for Free Field Theories,''
  JHEP {\bf 1204}, 074 (2012)
  doi:10.1007/JHEP04(2012)074
  [arXiv:1111.6290 [hep-th]].
  %%CITATION = doi:10.1007/JHEP04(2012)074;%%
  %44 citations counted in INSPIRE as of 06 Dec 2015
  
 %\cite{Fursaev:2012mp}
\bibitem{Fursaev:2012mp} 
  D.~V.~Fursaev,
  ``Entanglement Renyi Entropies in Conformal Field Theories and Holography,''
  JHEP {\bf 1205}, 080 (2012)
  doi:10.1007/JHEP05(2012)080
  [arXiv:1201.1702 [hep-th]].
  %%CITATION = doi:10.1007/JHEP05(2012)080;%%
  %27 citations counted in INSPIRE as of 06 Dec 2015
  
 %\cite{Dowker:2012rp}
\bibitem{Dowker:2012rp} 
  J.~S.~Dowker,
  ``Sphere Renyi entropies,''
  J.\ Phys.\ A {\bf 46}, 225401 (2013)
  doi:10.1088/1751-8113/46/22/225401
  [arXiv:1212.2098 [hep-th]].
  %%CITATION = doi:10.1088/1751-8113/46/22/225401;%%
  %7 citations counted in INSPIRE as of 06 Dec 2015
  
 %\cite{Huang:2014pfa}
\bibitem{Huang:2014pfa} 
  K.~W.~Huang,
  ``Central Charge and Entangled Gauge Fields,''
  Phys.\ Rev.\ D {\bf 92}, no. 2, 025010 (2015)
  doi:10.1103/PhysRevD.92.025010
  [arXiv:1412.2730 [hep-th]].
  %%CITATION = doi:10.1103/PhysRevD.92.025010;%%
  %11 citations counted in INSPIRE as of 06 Dec 2015
  
 %\cite{Eling:2013aqa}
\bibitem{Eling:2013aqa} 
  C.~Eling, Y.~Oz and S.~Theisen,
  ``Entanglement and Thermal Entropy of Gauge Fields,''
  JHEP {\bf 1311}, 019 (2013)
  doi:10.1007/JHEP11(2013)019
  [arXiv:1308.4964 [hep-th]].
  %%CITATION = doi:10.1007/JHEP11(2013)019;%%
  %9 citations counted in INSPIRE as of 06 Dec 2015
  
%\cite{Safdi:2012sn}
\bibitem{Safdi:2012sn} 
  B.~R.~Safdi,
  ``Exact and Numerical Results on Entanglement Entropy in (5+1)-Dimensional CFT,''
  JHEP {\bf 1212}, 005 (2012)
  doi:10.1007/JHEP12(2012)005
  [arXiv:1206.5025 [hep-th]].
  %%CITATION = doi:10.1007/JHEP12(2012)005;%%
  %23 citations counted in INSPIRE as of 06 Dec 2015
  
  %\cite{Lu:1998nu}
\bibitem{Lu:1998nu} 
  H.~Lu, C.~N.~Pope and J.~Rahmfeld,
  ``A Construction of Killing spinors on S**n,''
  J.\ Math.\ Phys.\  {\bf 40}, 4518 (1999)
  doi:10.1063/1.532983
  [hep-th/9805151].
  %%CITATION = doi:10.1063/1.532983;%%
  %135 citations counted in INSPIRE as of 06 Dec 2015
  
 %\cite{Galante:2013wta}
\bibitem{Galante:2013wta} 
  D.~A.~Galante and R.~C.~Myers,
  ``Holographic Renyi entropies at finite coupling,''
  JHEP {\bf 1308}, 063 (2013)
  doi:10.1007/JHEP08(2013)063
  [arXiv:1305.7191 [hep-th]].
  %%CITATION = doi:10.1007/JHEP08(2013)063;%%
  %13 citations counted in INSPIRE as of 06 Dec 2015
  
 %\cite{Manvelyan:2000ef}
\bibitem{Manvelyan:2000ef} 
  R.~Manvelyan and A.~C.~Petkou,
  ``A Note on R currents and trace anomalies in the (2,0) tensor multiplet in d = 6 AdS / CFT correspondence,''
  Phys.\ Lett.\ B {\bf 483}, 264 (2000)
  doi:10.1016/S0370-2693(00)00568-2
  [hep-th/0003017].
  %%CITATION = doi:10.1016/S0370-2693(00)00568-2;%%
  %8 citations counted in INSPIRE as of 16 Nov 2015
  
%\cite{Assel:2015nca}
\bibitem{Assel:2015nca} 
  B.~Assel, D.~Cassani, L.~Di Pietro, Z.~Komargodski, J.~Lorenzen and D.~Martelli,
  ``The Casimir Energy in Curved Space and its Supersymmetric Counterpart,''
  JHEP {\bf 1507}, 043 (2015)
  doi:10.1007/JHEP07(2015)043
  [arXiv:1503.05537 [hep-th]].
  %%CITATION = doi:10.1007/JHEP07(2015)043;%%
  %16 citations counted in INSPIRE as of 05 Dec 2015
  
 %\cite{Alday:2013lba}
\bibitem{Alday:2013lba} 
  L.~F.~Alday, D.~Martelli, P.~Richmond and J.~Sparks,
  ``Localization on Three-Manifolds,''
  JHEP {\bf 1310}, 095 (2013)
  doi:10.1007/JHEP10(2013)095
  [arXiv:1307.6848 [hep-th]].
  %%CITATION = doi:10.1007/JHEP10(2013)095;%%
  %46 citations counted in INSPIRE as of 01 Dec 2015
  
  %\cite{Closset:2013vra}
\bibitem{Closset:2013vra} 
  C.~Closset, T.~T.~Dumitrescu, G.~Festuccia and Z.~Komargodski,
  ``The Geometry of Supersymmetric Partition Functions,''
  JHEP {\bf 1401}, 124 (2014)
  doi:10.1007/JHEP01(2014)124
  [arXiv:1309.5876 [hep-th]].
  %%CITATION = doi:10.1007/JHEP01(2014)124;%%
  %61 citations counted in INSPIRE as of 01 Dec 2015
  
%\cite{Closset:2014uda}
\bibitem{Closset:2014uda} 
  C.~Closset, T.~T.~Dumitrescu, G.~Festuccia and Z.~Komargodski,
  ``From Rigid Supersymmetry to Twisted Holomorphic Theories,''
  Phys.\ Rev.\ D {\bf 90}, no. 8, 085006 (2014)
  doi:10.1103/PhysRevD.90.085006
  [arXiv:1407.2598 [hep-th]].
  %%CITATION = doi:10.1103/PhysRevD.90.085006;%%
  %15 citations counted in INSPIRE as of 01 Dec 2015
  
%\cite{Assel:2014paa}
\bibitem{Assel:2014paa} 
  B.~Assel, D.~Cassani and D.~Martelli,
  ``Localization on Hopf surfaces,''
  JHEP {\bf 1408}, 123 (2014)
  doi:10.1007/JHEP08(2014)123
  [arXiv:1405.5144 [hep-th]].
  %%CITATION = doi:10.1007/JHEP08(2014)123;%%
  %30 citations counted in INSPIRE as of 01 Dec 2015
  
%\cite{Bobev:2015kza}
\bibitem{Bobev:2015kza} 
  N.~Bobev, M.~Bullimore and H.~C.~Kim,
  ``Supersymmetric Casimir Energy and the Anomaly Polynomial,''
  JHEP {\bf 1509}, 142 (2015)
  doi:10.1007/JHEP09(2015)142
  [arXiv:1507.08553 [hep-th]].
  %%CITATION = doi:10.1007/JHEP09(2015)142;%%
  %3 citations counted in INSPIRE as of 16 Nov 2015
  
 %\cite{Bhattacharya:2008zy}
\bibitem{Bhattacharya:2008zy} 
  J.~Bhattacharya, S.~Bhattacharyya, S.~Minwalla and S.~Raju,
  ``Indices for Superconformal Field Theories in 3,5 and 6 Dimensions,''
  JHEP {\bf 0802}, 064 (2008)
  doi:10.1088/1126-6708/2008/02/064
  [arXiv:0801.1435 [hep-th]].
  %%CITATION = doi:10.1088/1126-6708/2008/02/064;%%
  %94 citations counted in INSPIRE as of 06 Dec 2015
  
 %\cite{Kim:2012qf}
\bibitem{Kim:2012qf} 
  H.~C.~Kim, J.~Kim and S.~Kim,
  ``Instantons on the 5-sphere and M5-branes,''
  arXiv:1211.0144 [hep-th].
  %%CITATION = ARXIV:1211.0144;%%
  %62 citations counted in INSPIRE as of 06 Dec 2015
  
 %\cite{Lockhart:2012vp}
\bibitem{Lockhart:2012vp} 
  G.~Lockhart and C.~Vafa,
  ``Superconformal Partition Functions and Non-perturbative Topological Strings,''
  arXiv:1210.5909 [hep-th].
  %%CITATION = ARXIV:1210.5909;%%
  %62 citations counted in INSPIRE as of 06 Dec 2015
  
%\cite{Cvetic:1999xp}
\bibitem{Cvetic:1999xp} 
  M.~Cvetic {\it et al.},
  ``Embedding AdS black holes in ten-dimensions and eleven-dimensions,''
  Nucl.\ Phys.\ B {\bf 558}, 96 (1999)
  doi:10.1016/S0550-3213(99)00419-8
  [hep-th/9903214].
  %%CITATION = doi:10.1016/S0550-3213(99)00419-8;%%
  %339 citations counted in INSPIRE as of 16 Nov 2015
  
\end{thebibliography}
\end{document}